# In-vivo 6D heart motion analysis for emerging self-powered cardiac implants

Milad Hasani[1], John Huber[2], Benedict Kjærgaard[3,4], Tomas Zaremba[3,5], Alireza Rezania[1*], Sam Riahi[3,5]

*[1] Department of Energy, Aalborg University, Aalborg, Denmark*
*[2] Department of Engineering Science, University of Oxford, Oxford, United Kingdom*
*[3] Department of Clinical Medicine, Aalborg University, Aalborg, Denmark*
*[4] Department of Cardiothoracic Surgery, Aalborg University Hospital, Aalborg, Denmark*
*[5] Department of Cardiology, Aalborg University Hospital, Aalborg, Denmark*

## Abstract

Self-powered intracardiac implant devices show great promise for future clinical applications due to their extended operational lifespan and the potential to reduce the need for high-risk repeat surgeries. This study investigates the feasibility of harvesting energy from cardiac motion through in vivo testing of intracardiac devices. Comprehensive three-dimensional translational and rotational cardiac motions are captured in a porcine model using a miniaturized 9-degree-of-freedom motion sensor implanted at six strategic epicardial sites. Kinematic criteria are developed to evaluate the energy harvesting potential of each implant site based on the available kinetic energy, acceleration, and jerk factors. The recorded heart motion signals are analyzed and applied to a conceptual energy harvester proposed to identify the optimal implant site. The results reveal that the left ventricular apex emerges as a preferable site for energy harvesting, particularly at moderate heart rates. These findings offer valuable insights into optimizing self-powered intracardiac implants, reducing dependency on battery replacements, and enhancing long-term patient safety.

---

## Introduction

Intracardiac implant devices enable a broad spectrum of functionalities crucial for

---

Corresponding author; Email address: alr@energy.aau.dk (Alireza Rezania)

cardiovascular health, ranging from precise electrical pacing and life-saving defibrillation to comprehensive diagnostic monitoring and advanced therapeutic applications. Intracardiac leadless pacemakers (ICLPs) have revolutionized cardiac pacing by eliminating the risks associated with traditional lead-based pacemakers, though early models were limited to single-chamber pacing. Advances in wireless synchronization now enable multi-chamber pacing, expanding their potential as a safer and less invasive alternative for a broader range of patients. The ICLPs rely on miniature integrated batteries with limited energy capacity. Replacing a pacemaker device (when the battery is depleted) needs surgical procedures that carry inherent risks and can be life-threatening for patients [1]. To address this limitation, researchers are exploring alternative energy sources that provide a continuous power supply and eliminate the need for high-risk replacement procedures. Energy harvesting from ambient environmental sources is a promising approach for this application. For instance, the kinetic energy of the human body and heart has been investigated as a power source for implantable cardiac devices, especially ICLPs [2,3].

Recent advances in energy harvesting have explored various mechanisms to eliminate battery replacement and extend intracardiac device longevity [4]. Electromagnetic induction-based harvesters leverage heart motion to drive a mass imbalance oscillation generator or oscillating magnet [5,6]. These designs generated up to 80 μW, sufficient for pacemaker function but requiring careful placement for optimal performance. The electromagnetic energy harvesters have been improved by a compact design for leadless cardiac pacemakers [7,8], consisting of a miniaturized oscillating magnet system that converts heart motion into electrical energy via electromagnetic induction. Unlike larger harvesters, these miniature designs are optimized for implantation inside leadless pacemakers. Moreover, a vascular turbine system has been explored [9], which harnesses blood flow to generate power. While this offers a continuous energy supply, thrombus formation and long-term biocompatibility remain concerns. Similarly, an intracardiac flow-based electromagnetic harvester [10] has been investigated to utilize blood flow in the right ventricular outflow tract, producing up to 82.64 μW, but requiring further in vivo validation.

Unlike electromagnetic energy harvesters, which contain permanent magnetic elements that interact with MRI scanning's magnetic field, piezoelectric energy harvesters offer better MRI compatibility because their materials are mainly unaffected by magnetic fields. An early study [11] on piezoelectric energy harvesting for pacemakers investigated a fan-folded intracardiac device, predicting its performance from heart motion. However, the experimental prototype was considerably larger than the proposed miniature capsule. Moreover, the long-term in vivo performance of an innovative hexa-fold piezoelectric energy harvester has recently been investigated for self-powered leadless pacemakers [12]. In another study [2], a PEH based on multiple spiral piezoelectric beams is investigated under one-dimensional motion normal to the epicardium (the outermost layer of the heart) measured at different sites by a laser sensor. The results have shown that the motion of the implant site can considerably affect the harvested energy level.

The heart muscle, or myocardium, contains fibers arranged in a helical pattern, causing the heart to twist during each contraction [13]. This twisting motion, known as ventricular torsion, enhances the efficiency of blood ejection. Therefore, beyond 3D translational movements, the twisting/rotational cardiac motion should be considered in the design of endocardial energy harvesters. However, previous research mainly focused on deriving 1D and 3D translational cardiac motions via MRI scanning [14], accelerometers [15–17], 1D laser measurements [2], image processing [18,19], echocardiography [20], and biplane videofluoroscopy and radiopaque beads [21]. The rotational motion measurement has been unmet in the previous studies.

This study aims to fill a gap in the literature by completely characterizing both 3D translational and rotational cardiac motion patterns at various epicardial sites and heart rates. In this regard, a 9-degree-of-freedom (DOF) motion sensor is implanted in six predefined epicardial points of a living pig heart to measure 3D translational and rotational movements at various heart rates. We use the measured heart motion to identify optimal implant sites and to guide energy-harvester design. Since evaluating the energy harvesting performance of each implant site depends on the energy harvester design, this study develops three semi-generic criteria based on available kinetic energy, acceleration, and jerk to consider an extensive range of energy harvester designs. The six observed sites are evaluated based on the defined criteria to find optimal implant sites for increasing energy harvesting performance. Finally, an endocardial energy harvester is considered to evaluate the efficiency of the developed semi-generic criteria.

This paper is organized as follows. Section 2, "Materials and Methods," details the experimental setup, including the motion sensor configuration, in-vivo measurement procedures, post-processing kinematic analysis of recorded heart motion, and the developed energy harvesting criteria. Section 3, "Results" presents the findings from applying these criteria and evaluates the performance of a conceptual endocardial energy harvester. Finally, Section 4 offers the "Discussions" summarizing the key findings and discussing future research directions.

## Materials and Methods

This section details the experimental setup and methodologies used to characterize heart motion for energy harvesting. This begins with a comprehensive description of the miniaturized 9-degree-of-freedom motion sensor, including its components, custom printed circuit board design, and encapsulation for in-vivo implantation. Following this, the in-vivo animal study procedures are outlined, covering the surgical approach, sensor calibration, and the strategic placement of the sensor at various epicardial sites in a porcine model to capture 3D translational and rotational heart movements across different heart rates. Finally, the section introduces the post-processing kinematic analysis, which transforms the raw sensor data into a fixed reference frame and defines the energy harvesting criteria based on kinetic energy, acceleration, and jerk to evaluate the potential of each implant site.

### Motion sensor configuration

The proposed sensor for heart motion measurement must provide accurate and comprehensive motion data. Its configuration includes an accelerometer to measure 3D linear acceleration, a gyroscope to capture the angular velocity of the implant site, and a 3D magnetometer to assess the strength of the local magnetic field. By integrating these measurements through a fusion algorithm, the sensor can accurately determine the 3D orientation of the implant site, ensuring precise motion tracking. The Bosch BNO055 inertial measurement unit (IMU) fulfills these requirements, making it an ideal choice for this application. This compact, advanced sensor is an all-in-one IMU that integrates the accelerometer, gyroscope, and magnetometer into a single device. It also includes a built-in microcontroller to process sensor fusion algorithms, delivering reliable 3D motion and orientation data. Its compact form factor makes it highly suitable for implantation in medical studies, where minimizing size and weight is crucial to prevent interference with natural heart motion. The sensor’s accuracy and bandwidth in fusion mode are detailed in Supplementary

Table S1.

The BNO055 sensor is integrated with additional electronic components on a printed circuit board (PCB) to function as a complete sensor system in this study. Considering the space-constrained environment of heart motion measurement, a customized double-sided PCB design is utilized to reduce the sensor system's overall dimensions significantly. The miniaturized double-sided PCB’s design with dimensions 13.4 × 6 × 3 mm is shown in Fig. 1-(a). This approach allows a more compact and lightweight sensor, minimizing potential tissue impact.

We implanted the miniaturized sensor into the epicardium during the animal test. Fig. 1-(b) shows the miniaturized sensor placed into a blue silicone package to prevent possible blood interference and facilitate suturing. The total mass of the motion sensor with its silicone package is 2 g, comparable to commercial ICLPs, and is sufficiently small to avoid impacting heart motion. The miniaturized sensor board requires four primary connections to function: ground (GND), power supply (VCC, typically 3.3V), and two serial communication lines (SCL and SDA) for I2C protocol. These connections are essential for providing power to the sensor and enabling data transfer between the microcontroller and the BNO055 sensor. In this research, the Arduino Due is used as the microcontroller to interface with the BNO055 sensor.

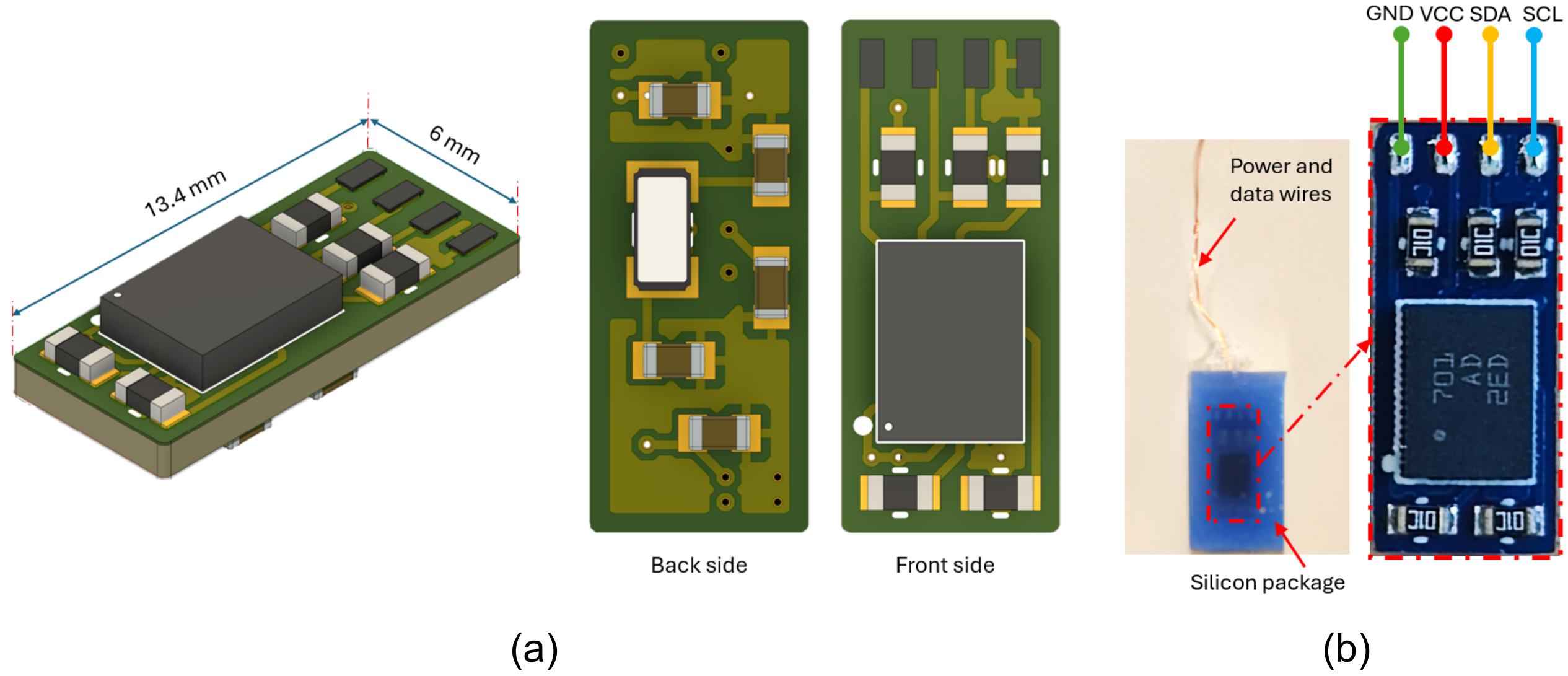

Fig. 1-(a) The miniaturized motion sensor with double-sided PCB design, (b) the miniaturized sensor encapsulated by a silicon package (with a blue color), and a zoom-in view of the actual sensor.

**In-vivo measurements**

We conducted an in-vivo animal study in a porcine model to measure heart motion. We utilized the porcine model as it is a standard analogue for human cardiac research, primarily due to its high degree of anatomical and physiological similarity to the human heart. We performed open-chest surgery to expose the heart and implant the sensor at different epicardial points and heart rates. We chose the locations for their relevance to typical leadless-pacemaker implantation sites. This study was approved by the Danish Animal Experiments Inspectorate, license no 2021-15-0201-00882. Detailed medical information of the animal test and procedural setup is provided in the Supplementary section S2.

The animal underwent an open-chest surgery to expose the heart and securely implant the motion sensor on different epicardial sites. Before sensor implantation, the motion sensor was carefully calibrated according to the manufacturer's instructions. Precise data collection was ensured by performing zero-bias correction to eliminate sensor drift, static and dynamic calibration to validate acceleration and angular velocity measurements, and magnetometer calibration to mitigate potential interference from the surgical environment. The operation of the sensor is also sensitive to electrostatic charge; therefore, to eliminate any interference during the animal test, all personnel used an earth wire. All measurements were taken from the same animal, during a single experiment.

We sutured the motion sensor (in a biocompatible silicone package) to six specific epicardial locations to measure 3D translational and rotational heart motion. These sites were selected based on their relevance to potential pacemaker implantation. During this animal test, a temporary pacemaker device was introduced through the venous catheter and connected to the heart to control heart rate. This temporary pacemaker allows controlled pacing at different specific rates, ensuring motion data is captured across various physiological conditions. The observed epicardial sites in this investigation are reported in Table 1. Moreover, Fig. 2. illustrates the implant sites over the epicardium with the motion sensor implanted at position 6. The shown relative distances between the implant sites are measured from a 2D projection to document the position at each acquisition, which may differ slightly from the true 3D separations. At implantation, the sensor's local axes were aligned to cardiac directions: Z is oriented perpendicular to the local epicardial surface (outward); Y was aligned base-to-apex; X is orthogonal to Y and Z, approximating the circumferential/torsional direction.

After implanting the motion sensor, we recorded real-time 3D translational accelerations, rotational velocities, and orientations at 50–70 Hz. We captured high-resolution cardiac motion data across a wide heart rate range (80–130 bpm), with over 20 cardiac cycles acquired per heart rate. Recorded data was transmitted via data cable to an Arduino Due board and subsequently to a computer for post-processing analysis.

Table 1- The measured motion cases at different epicardial implant sites and heart rates (HR) by sensor

| Positions | Placement |
|---|---|
| 1 | Mid-Septum right |
| 2 | Right ventricle – outflow tract |
| 3 | Basal lateral |
| 4 | High septum |
| 5 | Mid-anterior right ventricle |
| 6 | Apex, left ventricle |

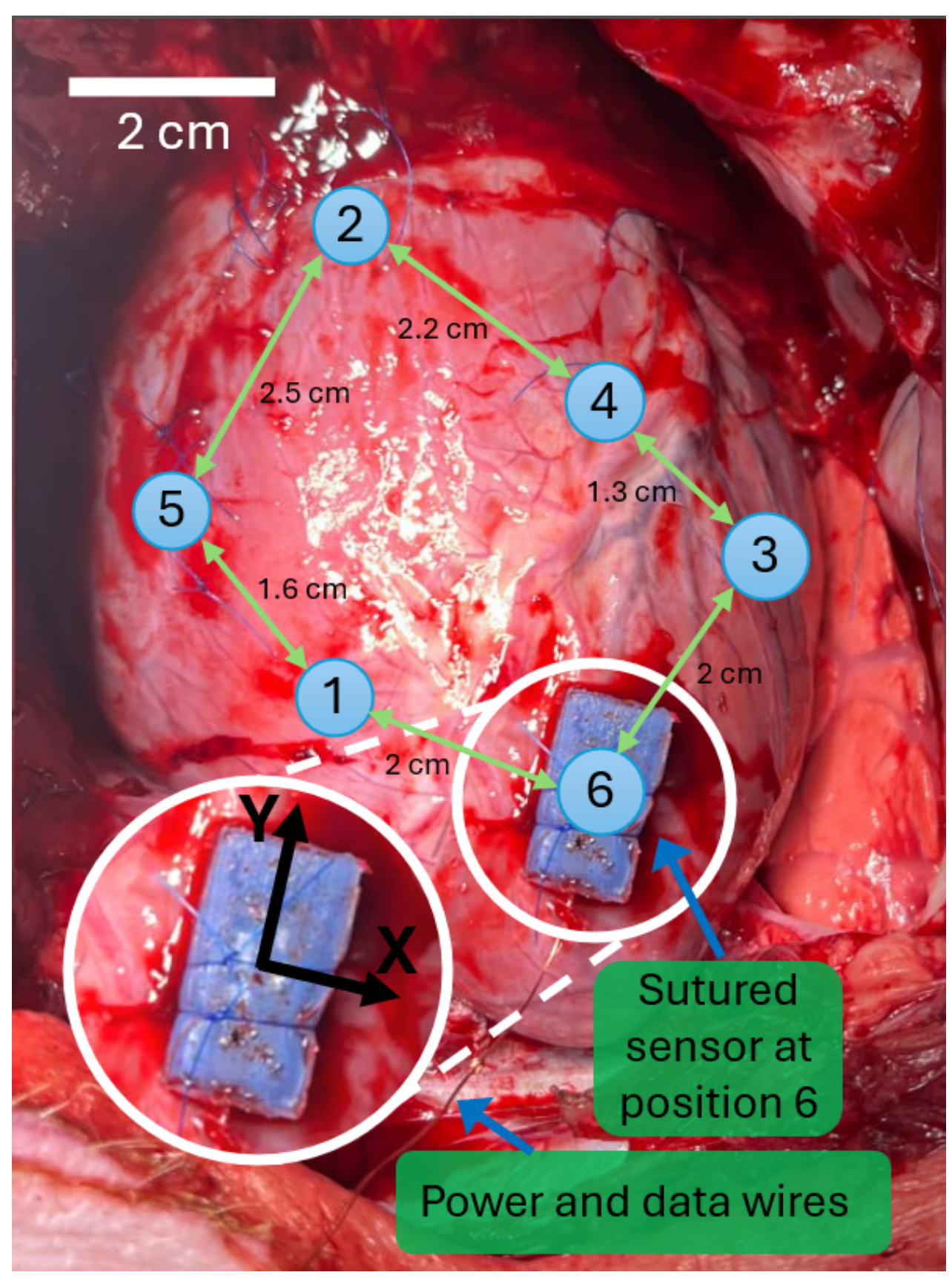

Fig. 2- The observed epicardial implant sites (by motion sensor), and the implanted sensor by suturing at position 6, along with the local motion sensor's coordinate system. The annotated distances reflect relative placement on the 2D view.

## Post-processing motion analysis

The motion sensor measures the 3D linear acceleration and angular velocity in the sensor's body frame, which follows the orientation of the sensor. It is required to transform measured data from the body frame to the fixed reference frame because the operation of proposed energy harvesters relates to inertial force that should be analyzed in the fixed reference coordinate system. Fig. 3-(a) shows the motion sensor implanted over the epicardium with an offset vector $\vec{r}$ from the heart surface. In supplementary section S3, a kinematic analysis is used to derive the corresponding base excitation that would be experienced by a self-powered ICLP through the fixation mechanism shown in Fig. 3-(b). As a result, the periodic base excitation, consisting of linear acceleration vector ($\overrightarrow{a'}_{base}$) and rotational velocity vector ($\overrightarrow{\Omega'}$) are computed.

It is supposed that the dimensions of the proposed self-powered ICLP are similar to current commercial pacemaker devices. For instance, the model of Abbott Aveir™ VR is characterized by a length of 38 mm and a diameter of 6.5 mm [2]. Usually, the battery occupies 60-70% of the pacemaker i.e., a cylindrical volume with a length 24.7 mm and diameter 6.5 mm. In the self-powered ICLP concept design, the battery volume can be allocated to the PEH part, as shown in Fig. 3-(b). Applying the derived motion vectors ($\overrightarrow{a'}_{base}$ and $\overrightarrow{\Omega'}$) to the base of ICLP create a time-dependent and position-dependent distribution of velocity and acceleration over the PEH volume, which should be analyzed for energy harvesting evaluation. The motion of each point within the capsule volume can be calculated either analytically or numerically by methods such as FEM. This research used a time-dependent model in COMSOL Multiphysics for kinematic analysis.

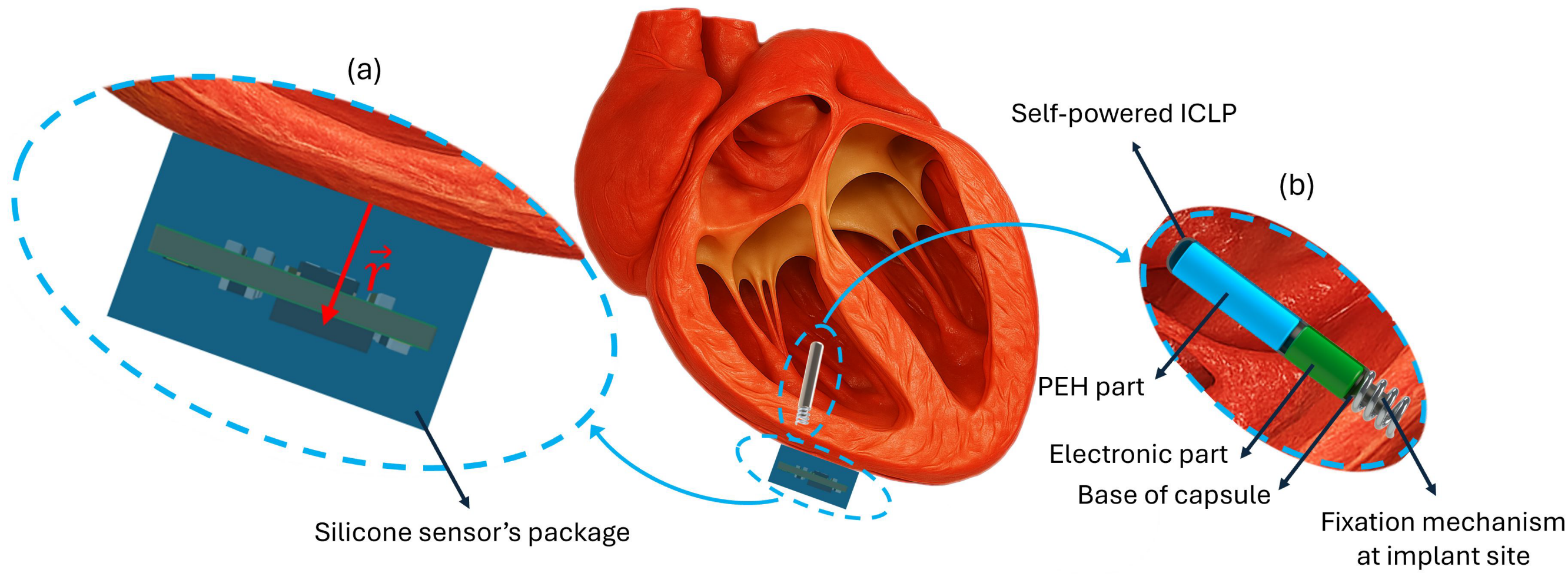

Fig. 3-The cross-sectional view of the heart, (a) showing the distance offset vector $\vec{r}$ between sensor and heart surface, (b) the cross-sectional view of an ICLP with a cylindrical capsule package over the endocardium.

The following analysis aims to identify optimal implant locations across varying heart rates based on energy harvesting potential. The optimal implant site(s) may vary depending on the specific energy harvester's design because the energy harvesting potential of each implant site depends on the energy harvester's properties. It is thus desirable to establish energy harvesting criteria that facilitate examining each implant site’s potential based on a broad range of energy harvester configurations.

### Energy harvesting criteria

In this study, several criteria are defined to encompass essential energy harvesting factors for a wide range of energy harvester designs. Specifically, three groups of criteria are developed based on (I) available kinetic energy (velocity), (II) acceleration, and (III) jerk, or acceleration derivative. All measured cases are evaluated against these criteria to predict each location’s energy harvesting potential at various heart rates. Finally, a specific energy harvester design is considered to examine the validity of the defined criteria.

### Criteria group I: The available kinetic energy

PEHs convert the kinetic energy of motion to electrical energy, so the total kinetic energy within the available capsule volume, $\mathbb{V}$, over a cycle of motion can be considered an energy harvesting criterion. As the inertial mass of the energy harvester is independent of the motion, only the magnitude of velocity at each point in the capsule volume contributes to a criterion for the evaluation of motion. Therefore, the first criteria group considers the time-averaged and volume-averaged velocity component magnitudes within $\mathbb{V}$ during one motion cycle, which correlates to the kinetic energy available for energy harvesting. This criteria group is expressed based on velocity components $v_x, v_y, v_z$ as:

$$Cr_{Ix} = \frac{1}{T}\int_0^T \frac{1}{\mathbb{V}} \iiint_{\mathbb{V}} |v_x(x,y,z,t)| \, d\mathbb{V} \, dt \qquad , [m/s] \tag{1}$$

with corresponding definitions for $Cr_{Iy}$ and $Cr_{Iz}$. Then, using the velocity magnitude:

$$Cr_I = \frac{1}{T}\int_0^T \frac{1}{\mathbb{V}} \iiint_{\mathbb{V}} \sqrt{v_x^2 + v_y^2 + v_z^2} \; d\mathbb{V} dt \qquad , [m/s] \tag{2}$$

where $T$ represents the duration of a cardiac cycle.

## Criteria group II: The available acceleration

In addition to the available kinetic energy (considered in the first criteria group), the available acceleration, and hence inertial force, is important in energy harvesting. Therefore, a second criteria group focuses on acceleration available within the PEH volume, which represents inertial loading potential. The time- and volume-averaged acceleration for *x*-component over a cycle of motion are given by:

$$Cr_{IIx} = \frac{1}{T}\int_{0}^{T} \frac{1}{\mathbb{V}} \iiint_{\mathbb{V}} |a_x(x,y,z,t)| \, d\mathbb{V}dt \quad , [m/s^2] \tag{3}$$

with corresponding definitions for $Cr_{IIy}$ and $Cr_{IIz}$. Then, the overall acceleration magnitude:

$$Cr_{II} = \frac{1}{T}\int_{0}^{T} \frac{1}{\mathbb{V}} \iiint_{\mathbb{V}} \sqrt{a_x^2 + a_y^2 + a_z^2} \, d\mathbb{V}dt \quad , [m/s^2] \tag{4}$$

## Criteria group III: The available jerk

PEHs convert an applied mechanical strain or stress to electrical output. In inertial systems, this is dependent on base acceleration. However, constant acceleration leads to a constant piezoelectric charge in such systems, whereas charge flow is required for the extraction of electrical energy.  Thus, in addition to the level of base acceleration, the time-varying acceleration is crucial for energy harvesting since constant stress or strain in the piezoelectric element does not lead to electrical power output [22,23]. In this way, the time derivative of acceleration, known as “jerk”, can be considered an effective factor in energy harvesting analysis, reflecting the dynamic variation necessary to induce charge flow in piezoelectric elements. Thus, the third criteria group ($Cr_{IIIx}, Cr_{IIIy}, Cr_{IIIz}, Cr_{III}$) examines the time-averaged and volume averaged available jerk within the PEH volume during one cardiac cycle as follows:

$$Cr_{IIIx} = \frac{1}{T}\int_{0}^{T} \frac{1}{\mathbb{V}} \iiint_{\mathbb{V}} |j_x(x,y,z,t)| \, d\mathbb{V}\,dt \quad , [m/s^3] \tag{5}$$

where $j_x = \frac{d}{dt} a_x$, with corresponding definitions for $Cr_{IIIy}$ and $Cr_{IIIz}$. Then, considering the overall jerk magnitude:

$$Cr_{III} = \frac{1}{T} \int_0^T \frac{1}{\mathbb{V}} \iiint_{\mathbb{V}} \sqrt{j_x^2 + j_y^2 + j_z^2} \, d\mathbb{V} \, dt \quad , [m/s^3] \tag{6}$$

In the defined criteria groups, the effect of both translational and rotational motions was considered. To estimate the contribution level of the 3D cardiac rotational motion in the criteria values, an expression $f_\Omega$ is defined to compute the fraction of each criterion's value arising due to rotational motion. Thus, for any of the assessment criteria $Cr_*$ the corresponding value of the same criterion evaluated without considering rotational motion $Cr_{*-\Omega}$ is computed and $f_\Omega$ is calculated by:

$$f_\Omega = \frac{|Cr_* - Cr_{*-\Omega}|}{Cr_*} \times 100\ \% \tag{7}$$

## Endocardial energy harvesting

This section studies an endocardial energy harvester under heart motion at the measured sites. Moreover, the effectiveness of the defined energy harvesting criteria is evaluated in this case study. The proposed energy harvester's design is based on a conventional cantilever piezoelectric beam configuration, as shown in Fig. 6. This design features a spatial distribution of PEH beams that enables the energy harvester to benefit from motion patterns throughout the cylinder volume. There are 25 piezoelectric beams with specific label numbers arranged in the proposed cylindrical space for the PEH part. Each piezoelectric beam consists of a bottom substrate layer (silicon), an upper piezoelectric layer (PZT-5H), and a tip mass (0.2 mg). The thickness of each piezoelectric and substrate layer is 20 µm. Each piezoelectric beam is connected to a separate electrical resistance ($R_L$) to measure the maximum energy extracted by the individual piezoelectric beams. The distributed piezoelectric beams in this design facilitate the investigation of energy harvesting levels at various locations within the energy harvester's volume.

In the literature [22,23], an analytical model has been developed to predict the response of cantilever piezoelectric energy harvesters (with bending motion) subjected to transverse base excitation (in the $z$-direction). In this study, the $z$-direction is aligned with the normal axis of the heart surface. The voltage generated under a harmonic transverse excitation with an amplitude $W_0$ and frequency $\omega$ ($w_b(t) = W_0 e^{j\omega t}$) is given by:

$$V(t) = \frac{\sum_{i=1}^{\infty} \frac{-j\omega F_i \theta_i}{-\omega^2 + 2j\xi_i \omega_i \omega + \omega_i^2}}{j\omega C_p + \frac{1}{R_L} + \sum_{i=1}^{\infty} \frac{j\omega \theta_i^2}{-\omega^2 + 2j\xi_i \omega_i \omega + \omega_i^2}} e^{j\omega t} \quad , [V] \tag{8}$$

where parameters $\theta_i$ and $\xi_i$ represent the equivalent modal electromechanical coupling and damping ratio, respectively. Moreover, the parameters $\omega_i$, $C_p$, and $R_\mathrm{L}$ are the ith undamped natural frequency, piezoelectric capacitance, and the connected electrical load's resistance, respectively. The $F_i$ represent the equivalent modal force of the proposed piezoelectric beam with length $L$, width $b$, and tip mass $M_t$, which can be expressed as

$$F_i = -\left[ (\rho_b h_b + 2\rho_p h_p) \int_0^L b(x) \phi_i(x)\, dx + M_t \phi_i(L) \right] W_0 \omega^2 = -m W_0 \omega^2 \tag{9}$$

The constants $\rho_b$, $\rho_p$, $h_b$, and $h_p$ denote the density and thickness of the substrate and piezoelectric layers, respectively. The function $\phi_i(x)$ is ith mode shape. Consequently, the fundamental frequency of the proposed piezoelectric beams is significantly higher than the harmonics present in heart motion ($\omega_1 \gg \omega$) due to their low length. In this case, considering only the fundamental mode is sufficient; the higher-order resonant modes can be neglected. Thus, Eq. (8) simplifies to:

$$V(t) = \frac{-jm\theta_1}{\omega_1^2 \left( j\omega C_p + \frac{1}{R_L} + \frac{j\omega \theta_1^2}{\omega_1^2} \right)} j_z(t) \quad , [V] \tag{10}$$

where the variable $j_z(t)$ is the applied transverse jerk given by

$$j_z(t) = -W_0 \omega^3 e^{j\omega t} \quad , [m/s^3] \tag{11}$$

The voltage output $V(t)$ is proportional to the transverse jerk $j_z(t)$ for each individual harmonic of the motion. Moreover, it can be shown that the effect of other jerk components $j_x(t)$ and $j_y(t)$ is negligible for a piezoelectric cantilever beam lying in the x-y plane. Large rotations between the body frame and fixed frame can produce significant jerk components $j_x(t)$ and $j_y(t)$. However, due to the limited rotation of the heart surface, the effects of $j_x(t)$ and $j_y(t)$ are expected to be negligible for the proposed energy harvester in this study.

The instantaneous generated power of the *i*th piezoelectric beam under a specific heart motion excitation is

$$P_i\,(t) = \frac{V_i^2(t)}{R_L} \quad ,[W] \tag{12}$$

where $V_i(t)$ is the voltage of the *i*th piezoelectric beam. The overall harvested power under the considered heart motion excitation is computed by summing over all beams:

$$p_{total}(t) = \sum_{i=1}^{25} P_i(t) \quad ,[W] \tag{13}$$

Normalizing with respect to the peak instantaneous power gives:

$$P_{total}(t) = \frac{p_{total}(t)}{\max(p_{total})} \tag{14}$$

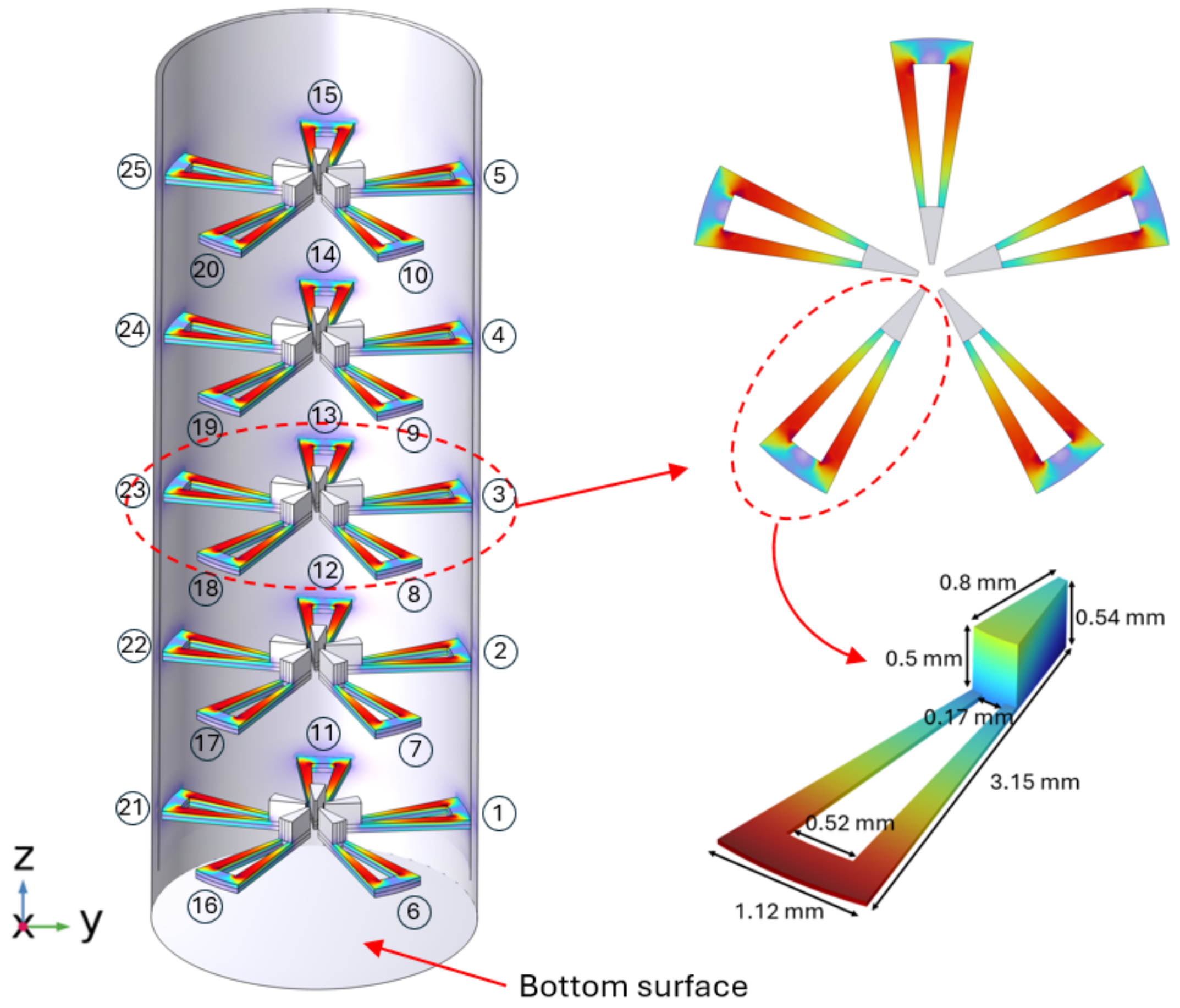

Fig. 4- The proposed endocardial energy harvester design, consisting of 25 individual piezoelectric beams, subjected to heart motion applied through the bottom surface; the color contour maps the von Mises stress distribution.

A finite element model of the energy harvester (illustrated in Fig. 4) is used to examine electrical output with measured heart motions as input. It is of interest to compare the normalized instantaneous power $P_{total,m}(t)$ to the time dependent volume averages of velocity, acceleration, and jerk given by:

$$\overline{v_x}(t) = \frac{1}{\mathbb{V}} \iiint_{\mathbb{V}} |v_x(x,y,z,t)|\, d\mathbb{V} \qquad , [m/s] \tag{15}$$

along with corresponding expressions for the *y* and *z* directions, for accelerations ($\overline{a_x}(t)$ etc.) and for jerk ($\overline{j_x}(t)$ etc.).

This finite element model is conducted in COMSOL Multiphysics using the solid mechanics, electrostatics, and electrical circuits modules. Each piezoelectric beam has an electrical connection to a separate resistance $R_L$=50 kΩ through individual voltage terminals. The piezoelectric beams are connected to the cylinder through a fixed constraint. Thin electrically conductive layers are applied to both the upper and bottom surfaces of the piezoelectric layers to collect and transport electrical charges generated within the material subjected to mechanical stress. The isotropic material properties of density, Young's modulus, and Poisson ratio for the silicon layer are 2329 kg/m$^3$, 170 GPa, and 0.28, respectively. Moreover, the material properties of the piezoelectric layer (PZT-5H) are presented in the Appendix. The isotropic loss factor $\eta_s = 0.01$ was considered as a mechanical damping.

## Results

This section presents a comprehensive analysis of the energy harvesting potential at various epicardial implant sites and heart rates, using the criteria defined in the preceding sections. The evaluation is based on three groups of criteria: available kinetic energy (velocity), acceleration, and jerk. The results from these criteria are discussed, highlighting how different implant sites perform under varying physiological conditions. Additionally, the influence of rotational cardiac motion on the criteria values is assessed. Finally, an endocardial energy harvester design is analyzed to validate the effectiveness of the established criteria in predicting power output from heart motion.

### Analysis of energy harvesting potential of implant sites

The heart motion across all implant sites (as stated in Table 1) is comprehensively analyzed using the defined energy harvesting criteria at different heart rates to provide a statistical assessment of measurement repeatability and uncertainty. The 20 cardiac cycles recorded for each case were grouped into heart-rate bins for each of the six implant positions. The detailed results are provided in Supplementary Section S4.

To compare the energy-harvesting potential of implant sites, Fig. 5 summarizes the median values of the energy-harvesting criteria for each heart rate bin. The first, second, and third columns in this figure show the results for the first, second, and third groups of criteria, respectively. The most consistent trend across all 12 graphs is a positive correlation between complexity and heart rate. However, some criteria are not considerably affected by heart rate, such as $Cr_{Iy}$.

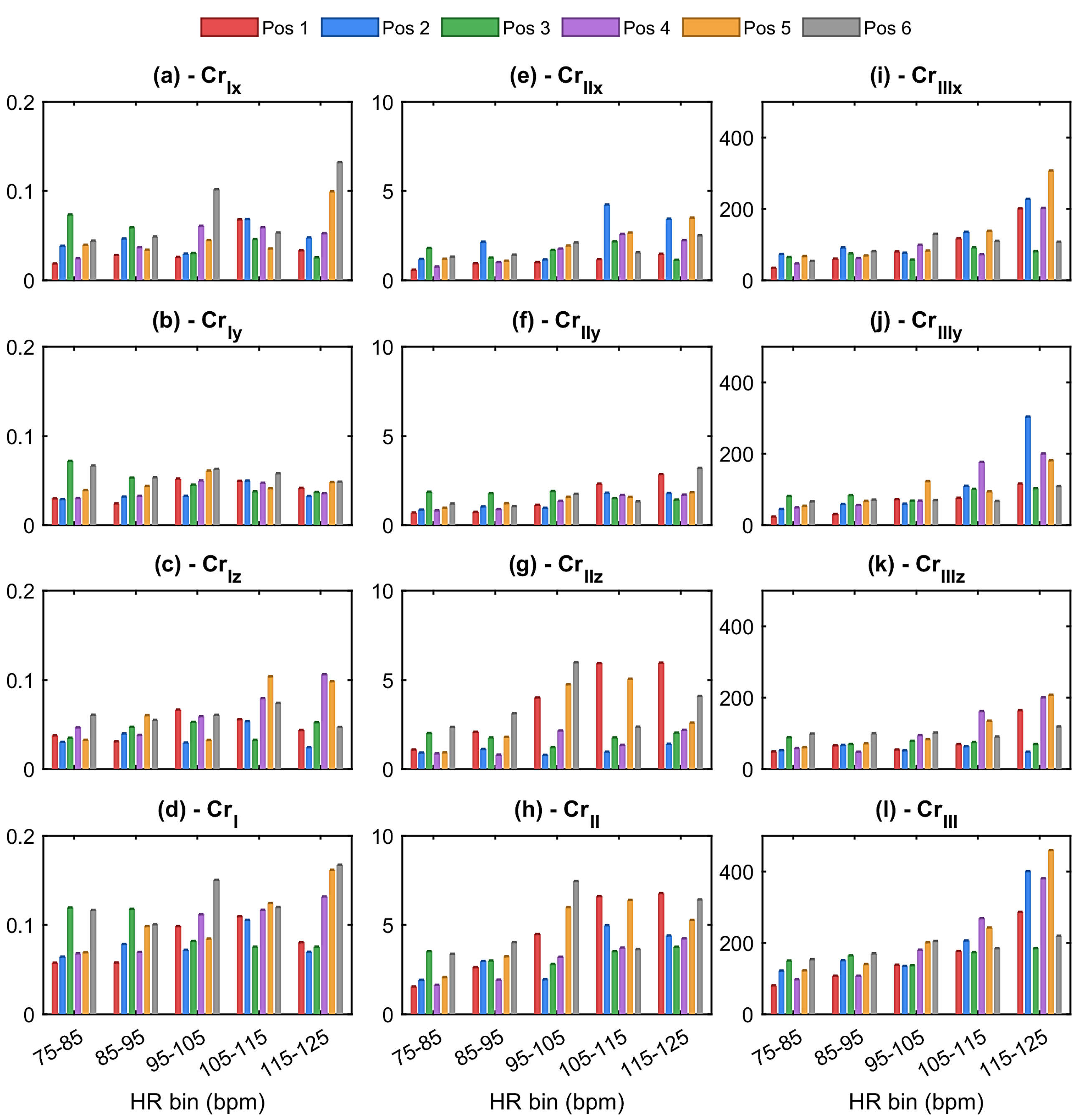

*The figures in the first, second, and third columns are plotted in SI units: m/s, m/s$^2$, and m/s$^3$, respectively.

Fig. 5- The median values of criteria at different implant sites and heart-rate bins; kinetic energy criteria: (a) $Cr_{Ix}$, (b) $Cr_{Iy}$, (c) $Cr_{Iz}$, (d) $Cr_{I}$, acceleration criteria: (e) $Cr_{IIx}$, (f) $Cr_{IIy}$, (g) $Cr_{IIz}$, (h) $Cr_{II}$, and jerk criteria: (i) $Cr_{IIIx}$, (j) $Cr_{IIIy}$, (k) $Cr_{IIIz}$, (l) $Cr_{III}$. Position 6 (grey) and Position 5 (orange) generally show the highest median values across all criteria. In contrast, Position 2 (blue) consistently shows the lowest motion levels.

Due to the high complexity and variations in Fig. 5, it is complicated to provide a comprehensive analysis or consistent patterns across all directions and implant sites. In this way, Table 2 summarizes the directional mean criteria values ($\overline{Cr}$) in all 12 subplots in Fig. 5 throughout all implant sites and heart-rate bins. As shown in the supplementary material and Fig. 2, the x-, y-, and z-axes of the sensor's coordinate system were aligned with the heart's torsional, longitudinal, and normal motions throughout all measurements. Table 2 indicates that the first and third criteria are maximal in the torsional direction, whereas the second criterion is maximal in the normal direction. Moreover, the averaged results show that the longitudinal direction has the lowest motion level; however, the third criteria group's result for implant site 2 (in Fig. 5) is maximal in the longitudinal direction.

Table 2- Directional average of the criteria values ($\overline{Cr}$) in Fig. 5 throughout all implant sites and heart-rate bins.

| **Criteria group** | **Criteria values in different directions** | | | |
|---|---|---|---|---|
| | **Torsional (x)** | **Longitudinal (y)** | **Normal (z)** | **Total** |
| First (m/s) | $\overline{Cr_{Ix}}$ = 0.051 | $\overline{Cr_{Iy}}$ = 0.044 | $\overline{Cr_{Iz}}$ = 0.048 | $\overline{Cr_{I}}$ = 0.10 |
| Second (m/s$^2$) | $\overline{Cr_{IIx}}$ = 1.830 | $\overline{Cr_{IIy}}$ = 1.56 | $\overline{Cr_{IIz}}$ = 2.46 | $\overline{Cr_{II}}$ 3.94 |
| Third (m/s$^3$) | $\overline{Cr_{IIIx}}$ =101.7 | $\overline{Cr_{IIIy}}$ = 93.6 | $\overline{Cr_{IIIz}}$ = 89.9 | $\overline{Cr_{III}}$ = 192.9 |

To facilitate a clearer, normalized comparison of the sites, the median criteria values presented in Figure 5 were converted into a scoring system for Table 3. For each three multidirectional criteria ($\overline{Cr_{I}}$, $\overline{Cr_{II}}$, $\overline{Cr_{III}}$) and each specific heart-rate bin, a position's score was calculated as the ratio of its median value to the sum of the median values from all six positions, then multiplied by 100. This score effectively represents each site's percentage contribution to the total motion captured across all sites for that specific criterion and heart rate. The highlighted cells within the table identify the optimal implant site for each criterion within each specific heart-rate (HR) range. This data illustrates that the best-performing location is not consistent, but rather changes depending on both the heart rate and the specific motion characteristic (velocity, acceleration, or jerk) being measured. For example, while Position 6 (Apex, left ventricle) and Position 5 (Mid-anterior right ventricle) frequently show high scores, Position 3 (Basal lateral) is optimal for acceleration and jerk criteria at lower heart rates, and Position 4 (High septum) is optimal for the jerk criterion in the 105-115 bpm range.

Table 3- The average of the criteria values of $\overline{Cr_I}$, $\overline{Cr_{II}}$, and $\overline{Cr_{III}}$ at each heart-rate bin and each implant site.

| Implant sites | Implant sites' performance score (0-100) by HR Range (BPM) | | | | | | | | | | | | | | |
|---|---|---|---|---|---|---|---|---|---|---|---|---|---|---|---|
| | 75-85 | | | 85-95 | | | 95-105 | | | 105-115 | | | 115-125 | | |
| | $\overline{Cr_I}$ | $\overline{Cr_{II}}$ | $\overline{Cr_{III}}$ | $\overline{Cr_I}$ | $\overline{Cr_{II}}$ | $\overline{Cr_{III}}$ | $\overline{Cr_I}$ | $\overline{Cr_{II}}$ | $\overline{Cr_{III}}$ | $\overline{Cr_I}$ | $\overline{Cr_{II}}$ | $\overline{Cr_{III}}$ | $\overline{Cr_I}$ | $\overline{Cr_{II}}$ | $\overline{Cr_{III}}$ |
| Pos 1 | 11.6 | 10.9 | 11.0 | 11.0 | 14.8 | 12.8 | 16.5 | 17.3 | 13.9 | 17.3 | 24.2 | 14.5 | 11.1 | 21.5 | 16.5 |
| Pos 2 | 13.0 | 13.6 | 16.8 | 15.0 | 16.7 | 18.0 | 12.0 | 7.5 | 13.5 | 14.0 | 12.6 | 14.0 | 14.6 | 15.8 | 11.9 |
| Pos 3 | 24.1 | 25.1 | 20.7 | 22.6 | 16.8 | 19.6 | 13.7 | 10.8 | 13.7 | 11.9 | 12.9 | 14.3 | 10.5 | 12.0 | 10.6 |
| Pos 4 | 13.7 | 11.6 | 13.4 | 13.3 | 10.8 | 12.8 | 18.7 | 12.4 | 18.1 | 18.4 | 13.6 | 22.1 | 18.2 | 13.5 | 21.9 |
| Pos 5 | 14.0 | 14.7 | 16.9 | 18.8 | 18.2 | 16.7 | 14.1 | 23.1 | 20.2 | 19.6 | 23.4 | 20.0 | 22.4 | 16.8 | 26.5 |
| Pos 6 | 23.6 | 24.1 | 21.2 | 19.2 | 22.7 | 20.2 | 25.1 | 28.8 | 20.6 | 18.9 | 13.3 | 15.2 | 23.2 | 20.4 | 12.6 |

## Contribution of rotational motion

To assess the contribution level of the 3D rotational motion in energy harvesting criteria, the average of $f_\Omega$ values (defined in Eq. (7) of some measurements with different heart rates at individual implant sites for each criterion are reported in Fig. 6. The results indicate that although the velocity, acceleration, and jerk components in the *x*- and *y*-directions are significantly affected by rotation motion, the components in the *z*-direction are only slightly changed by rotation. This arises because the rotation angles are relatively small (only a few degrees), and the offset between the epicardium and the sensor remains approximately in the *z*-direction throughout the cardiac cycle. Therefore, the criteria $Cr_{Iz}$, $Cr_{IIz}$, and $Cr_{IIIz}$ are not greatly influenced by rotational motion (as shown in Fig. 6).

The motion of the proposed energy harvester's volume (at different time instants) is analyzed subject to a measured cardiac motion. The resulting distribution of jerk magnitude over this volume is presented in Fig. 6-(b-e), along with velocity arrow lines under base excitation ($\overrightarrow{a'}_{base}(t)$ and $\overrightarrow{\Omega'}(t)$) with different views at time instants: (1) *t* = 3.2 ms, (2) *t* = 90 ms, and (3) *t* = 200 ms. The shown variation in jerk distribution illustrates the effect of rotational motion on the energy harvester's kinematics. The effect of rotational motion in instantaneous power generation is shown in Supplemental Figure S9.

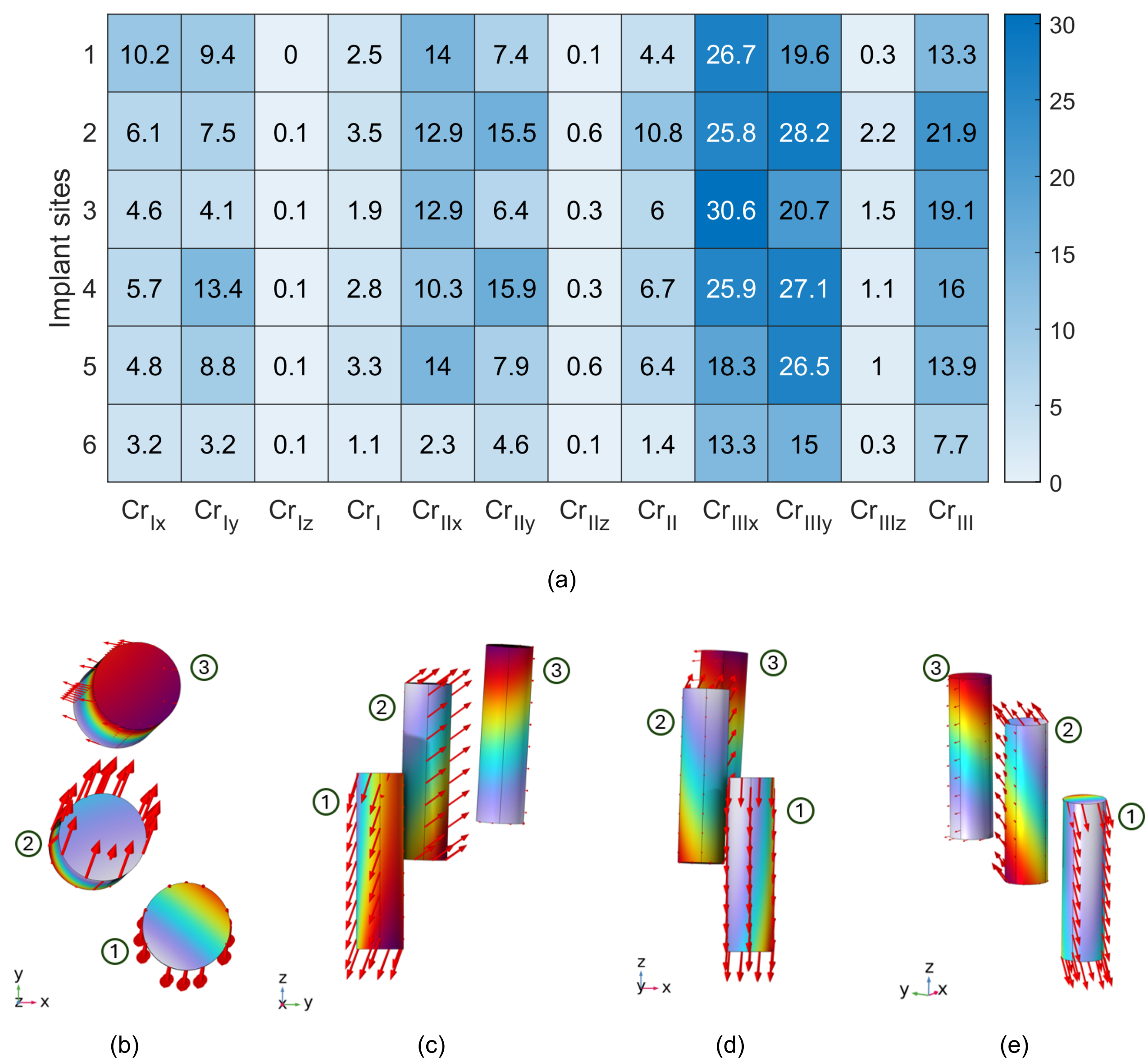

Fig. 6-(a) The effect of rotational motion ($f_{\Omega}$) in each criterion at different implant sites, the distribution of jerk magnitude within the energy harvester's cylindrical volume along with velocity arrow lines at different time instants ((1) t=3 ms, (2) t=90 ms, and (3) t=200 ms) in (b) x-y, (c) y-z,(d) z-x, and (e) perspective views.

## Performance of energy harvesting criteria

This section evaluates the performance of the defined criteria in predicting the energy-harvesting potential of implant sites. In this way, the output of the energy harvester shown in Fig. 4 is examined under a specific measured heart motion to determine which motion measure best correlates with instantaneous power generation. The results of the instantaneous volume average of velocity, acceleration, and jerk defined in Eq. (15) and corresponding expressions are presented in Fig. 7-(a-c), respectively. Fig. 7-(a) and Fig. 7-

(b) indicate that there is no clear correlation between instantaneous power and either velocity or acceleration. However, Fig. 7-(c) shows that the variation of instantaneous power aligns with the measure of available jerk $\overline{J_z}(t)$. For instance, the peaks of instantaneous power occur when $\overline{J_z}(t)$ reaches its peaks. However, due to the beam alignments for the proposed energy harvester design (in Fig. 4), there is no explicit correlation between instantaneous power and in-plane jerk components $\overline{J_x}(t)$ and $\overline{J_y}(t)$.

Eq. (10) and Eq. (12) suggest that the total generated power is proportional to the square of the instantaneous volume-averaged jerk magnitude,

$$P_{total}(t) \propto \overline{J_z}^2(t) \tag{16}$$

To evaluate this correlation, the squared transverse jerk normalized to its peak value is compared with the normalized power $\bar{P}_{total}(t)$ in Fig. 7-(d). The result demonstrates that the energy harvester's instantaneous power correlates very closely with the instantaneous $\overline{J_z}^2(t)$. Thus, it can be expected that the energy harvesting criterion $Cr_{IIIz}$ is of value in predicting the energy harvesting potential of different locations.

Some heart motion measurements at different implant sites and heart rates are considered, and the corresponding average power (during the cardiac cycle) is computed. The derived values are normalized to the lowest average power value, and the normalized values ($P_{avg}$) are shown in Fig. 8 and compared against the corresponding $Cr_{IIIz}^2$. The results indicate that there is a close correlation between the normalized average power $P_{avg}$ and the predicted values by $Cr_{IIIz}^2$. Thus, the expression in Eq. (16) can be extended to $P_{avg} \propto Cr_{IIIz}^2$ for this specific energy harvesting configuration.

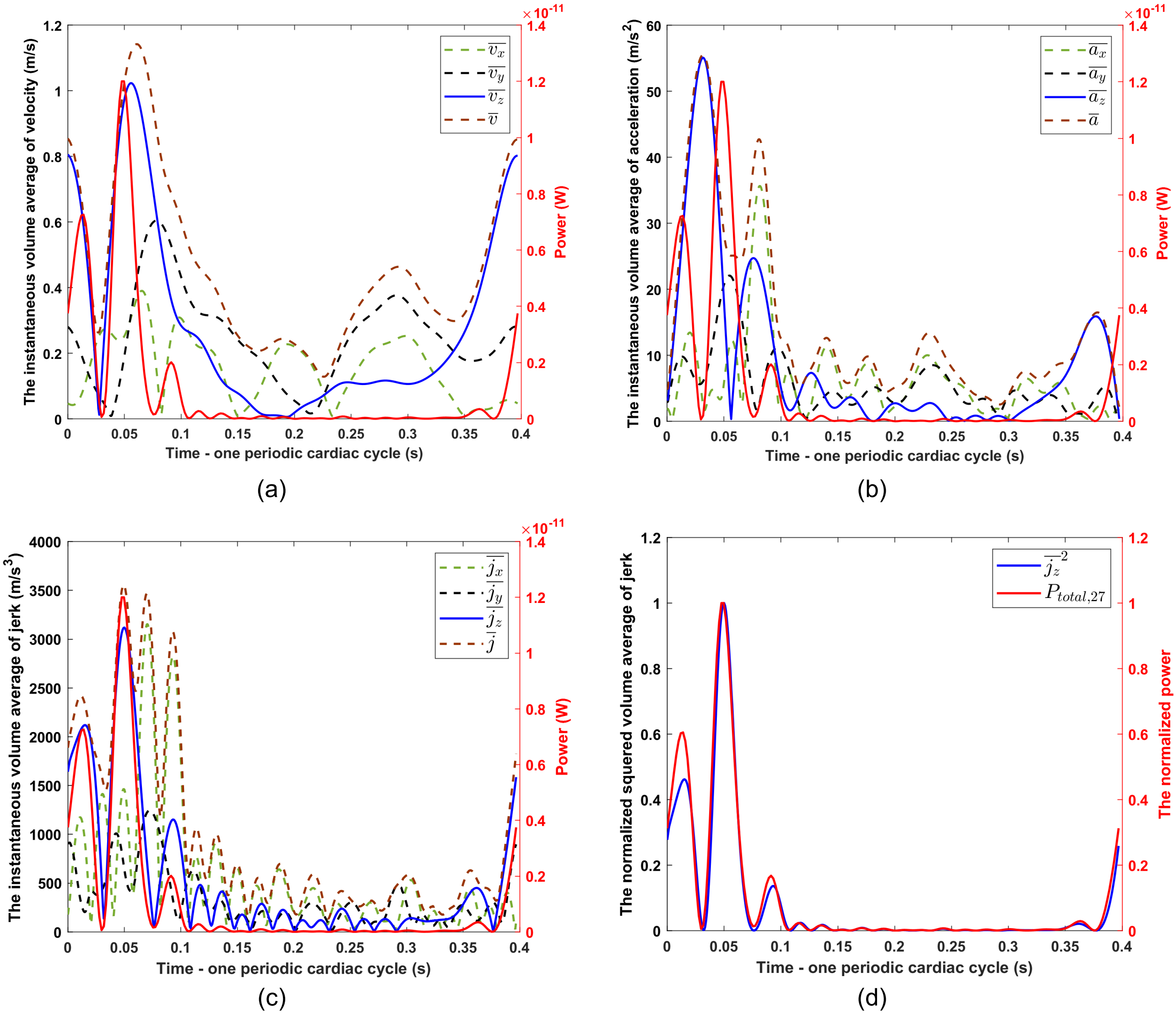

Fig. 7- The comparison of normalized whole power against  the (a) instantaneous volumetric absolute of velocity, (b) acceleration, (c) jerk, and (d) normalized squared transverse jerk $\overline{j_z}^2(t)$. A direct comparison shows that the normalized instantaneous power correlates very closely with the normalized squared transverse jerk ($\overline{j_z^2}$), validating $Cr_{IIIz}$ as a predictive criterion for this harvester design.

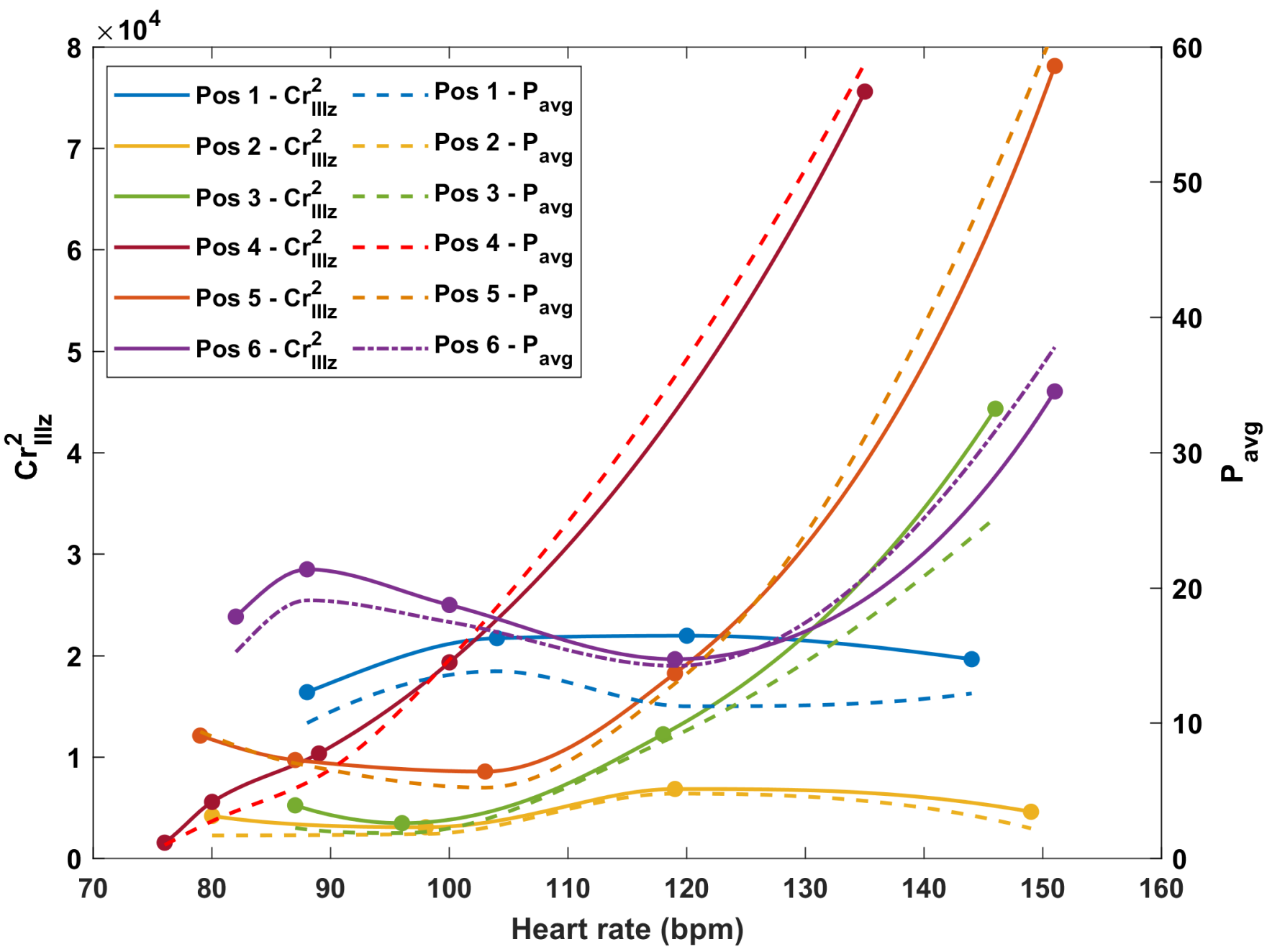

Fig. 8- The normalized average power $P_{avg}$ compared to $Cr_{IIIz}^2$ at different sites and heart rates, confirming that $Cr_{IIIz}$ is a valid and efficient proxy for site selection for this harvester type.

## Discussion

This result facilitates evaluating the relative energy harvesting potential of implant sites without requiring an electromechanical analysis of the energy harvester. It demonstrates that, for certain designs of inertial energy harvester, the energy harvesting potential of implant sites on the heart can be assessed by analyzing the motion to find the total available jerk rather than performing an electromechanical analysis of the energy harvester itself at individual implant sites and heart rates. Indeed, the defined criteria provide an efficient analytical framework for evaluating each implant site, significantly decreasing computational load and time.

The proposed energy harvester solely benefits from the jerk component normal to the heart surface ($j_z$); however, Fig. 7 shows that the level of other jerk components $j_x$ and $j_y$ is significant. Therefore, the best energy harvester design should be able to take advantage of multidirectional jerk. In this situation, the criterion $Cr_{III}$ can be considered for jerk-based energy harvesting designs.

Although the third criteria group accurately predicts the behavior of certain inertial energy harvester designs, wherein jerk is the dominant source of current flow, it should be emphasized that the appropriate criterion depends on the operating principle of the specific energy harvester. For instance, the available kinetic energy (determined by the first criteria group) may be the most effective criterion for endocardial energy harvester designs based on impact, as investigated in [12]. In this way, Table 4 provides the average of scores in Table 3 at each implant sites. This average values is used to establish a ranking of the implant sites based on energy harvesting potential estimated by each multidirectional criterion ($Cr_I$, $Cr_{II}$, and $Cr_{III}$) over the heart rate range 75-125 bpm.

Based on the analysis in Table 4, which provides the average scores for all criteria across the entire 75-125 bpm heart rate range, a clear ranking of the implant sites by energy harvesting potential can be established. The results from the different criteria groups are largely consistent in identifying the most and least suitable locations. Position 6 (Apex, left ventricle) achieves the highest overall score (61.8), ranking it as the most promising site for an energy harvesting implant. It is followed by Position 5 (Mid-anterior right ventricle), which ranked second with a score of 57.1. Conversely, Position 2 (Right ventricle - outflow tract) received the lowest overall score (41.8), indicating it is the least suitable location among those tested. The complete ranking, in descending order of suitability based on the overall score, is: Position 6, Position 5, Position 3 (Basal lateral), Position 4 (High septum), Position 1 (Mid-Septum right), and Position 2.

Table 4- The scores and ranking of the observed positions based on energy harvesting capability evaluated by multidirectional criteria.

| **Implant sites** | **Average score** | | | **Overall score** | **Ranking** |
|---|---|---|---|---|---|
| | $Cr_I$ | $Cr_{II}$ | $Cr_{III}$ | | |
| Pos 1 | 13.5 | 17.7 | 13.7 | 45.0 | **5** |
| Pos 2 | 13.7 | 13.2 | 14.8 | 41.8 | **6** |
| Pos 3 | 16.6 | 15.5 | 15.8 | 47.9 | **3** |
| Pos 4 | 16.5 | 12.4 | 17.7 | 46.5 | **4** |
| Pos 5 | 17.8 | 19.2 | 20.1 | 57.1 | **2** |
| Pos 6 | 22.0 | 21.9 | 18.0 | 61.8 | **1** |

* Pos 1: Mid-Septum right, Pos 2: Right ventricle – outflow tract, Pos 3: Basal lateral, Pos 4: High septum, Pos 5: Mid-anterior right ventricle, Pos 6: Apex, left ventricle.

In conclusion, this study presented a novel approach for characterizing in-vivo cardiac motion to optimize kinetic-based energy harvesting for intracardiac implants. A 9-DOF motion sensor was miniaturized and implanted over the epicardium in an animal model to comprehensively measure and analyze 3D translational and rotational heart motion across multiple epicardial implant sites and heart rates. Three criteria groups based on available kinetic energy (velocity), acceleration, and jerk, were developed to examine the energy harvesting performance of each implant site at different heart rate ranges. Moreover, a theoretical model of an endocardial energy harvester based on distributed cantilever piezoelectric beams was proposed in order to evaluate the energy harvesting output at various implant sites and heart rates. The results highlight the importance of site selection for maximizing energy harvesting potential, with the left ventricular apex emerging as a promising location for jerk-based energy harvesters. The proposed jerk-based criterion effectively predicts power output based on motion data, offering a computationally efficient way to assess sites for future implant optimization. These findings pave the way for the development of self-powered intracardiac implants (e.g., ICLPs), potentially eliminating the need for high-risk battery replacement surgeries and improving patient outcomes.

To further validate these findings and move toward clinical translation, our next step involves placing the sensor endocardially using a percutaneous, closed-chest approach. This will allow us to assess fluid-structure interactions (between blood flow and capsule) and effects from papillary muscles. Future studies will also involve multiple animals to assess inter-subject variability. Finally, to bridge the translational gap between our animal model and human applications, we plan to use non-invasive cine magnetic resonance imaging in human subjects. This will allow us to reconstruct and quantitatively compare human cardiac motion profiles against the fundamental patterns identified in this study, helping to validate whether our relative findings translate to human physiology.

## Declarations

Ethics approval: The Danish Animal Experiments Inspectorate approved this study, license no 2021-15-0201-00882.

Consent for publication: Not applicable.

Clinical trial number: Not applicable

Availability of data: Data are available on request.

Competing interest: The authors declare that they have no known competing financial interests or personal relationships that could have appeared to influence the work reported in this paper.

Funding: This work was supported by a research grant from the Danish Cardiovascular Academy, which is funded by the Novo Nordisk Foundation, grant number NNF20SA0067242, and the Danish Heart Foundation. Also, the Beta Health and Health Hub Foundations supported the work. The computational resources in work were partially funded by DeiC National HPC (DeiC-AAU-N2-2025122).

Author Contributions: Milad Hasani: conceptualization, methodology, investigation, writing – original draft, software, data curation, visualization. John E. Huber: methodology, investigation, methodology, writing—review & editing. Benedict Kjærgaard: investigation, writing—review & editing. Tomas Zaremba: writing—review & editing. Alireza Rezania: conceptualization, resources, writing—review & editing, supervision, and project administration. Sam Riahi: methodology, investigation, resources, writing—review & editing, supervision, and project administration.

Acknowledgments: Not applicable.

## Appendix

The material properties of the piezoelectric layer (PZT-5H) are as follows:

- Elasticity matrix: $$[c^E] = \begin{bmatrix} 127.2 & 80.2 & 84.7 & 0 & 0 & 0 \\ 80.2 & 127.2 & 84.7 & 0 & 0 & 0 \\ 84.7 & 84.7 & 117.4 & 0 & 0 & 0 \\ 0 & 0 & 0 & 23.0 & 0 & 0 \\ 0 & 0 & 0 & 0 & 23.0 & 0 \\ 0 & 0 & 0 & 0 & 0 & 23.47 \end{bmatrix} GPa$$

- Coupling matrix: $$[e] = \begin{bmatrix} 0 & 0 & 0 & 0 & 17.03 & 0 \\ 0 & 0 & 0 & 17.03 & 0 & 0 \\ -6.62 & -6.62 & 23.24 & 0 & 0 & 0 \end{bmatrix} C/m^2$$

- Permittivity: $$[\varepsilon] = \begin{bmatrix} 1704.4 & 0 & 0 \\ 0 & 1704.4 & 0 \\ 0 & 0 & 1433.6 \end{bmatrix} \varepsilon_0, \qquad \varepsilon_0 = 8.854 \times 10^{-12} \quad F/m$$

# Supporting Information

## In-vivo 6D heart motion analysis for self-powered intracardiac implants development

Milad Hasani[1], John Huber[2], Benedict Kjærgaard[3,4], Tomas Zaremba[3,5], Alireza Rezania[1*], Sam Riahi[3,5]

[1] *AAU Energy, Aalborg University, Aalborg, Denmark*

[2] *Department of Engineering Science, University of Oxford, Parks Rd, Oxford, OX1 3PJ United Kingdom*

[3] *Department of Clinical Medicine, Aalborg University, Aalborg, Denmark*

[4] *Department of Cardiothoracic Surgery, Aalborg University Hospital, Aalborg, Denmark*

[5] *Department of Cardiology, Aalborg University Hospital, Aalborg, Denmark*

## S1. BNO055 fusion mode sensor characteristics

Table S1 outlines the key operating limits, bandwidths, resolutions, noise characteristics, and typical offsets of the accelerometer and gyroscope employed in fusion mode. These specifications define the baseline accuracy and dynamic response of the BNO055 sensor.

Table S1- Performance Specifications of accelerometer and gyroscope in Fusion Mode

| **Parameter** | **Accelerometer** | **Gyroscope** |
|---|---|---|
| Range | ±4g | ±2000 °/s |
| Bandwidth | 62.5 Hz | 32 Hz |
| Resolution | 14-bit | 16-bit |
| Noise Density | ~150 µg/√Hz | ~0.014 °/s/√Hz |
| Offset (Typical) | ±80 mg | ±1 °/s |

The noise and filtering relied on both the BNO055 sensor's built-in hardware capabilities and our post-processing software. The sensor itself has built-in frequency filters and an internal fusion algorithm that inherently reduces noise by blending sensor data. In post-processing, we also applied a software low-pass filter to remove any high-frequency noise above 60 Hz.

## S2. Detailed medical information of in-vivo experimentation

A female Danish Landrace-Yorkshire pig of approximately 50 kg was used for the experiment. Anesthesia was induced with Zoletil 10 ml intramuscular, a mixture of Tiletamine 8.3 mg/ml, Zolazepam 8.3 mg/ml, Butorphanol 1.7 mg/ml, and Xylazine 8.3 mg/ml. After a central venous line was obtained anesthesia was maintained with Propofol and Fentanyl delivered by infusion pumps and adjusted according to the animals' reactions. The animal was intubated for ventilation with a Dräger Primus ventilator (Dräger Medical Deutschland GmbH, Lübeck, Germany). Tidal volume was 10 ml/kg with an end expiratory pressure of 5 kPa; respiratory rate was adjusted according to an end tidal carbon dioxide concentration of 4-6 kPa.

A 6 French arterial catheter was inserted into a femoral artery for continuous blood pressure monitoring and a 10 Fr catheter was put into the left jugular vein for insertion of pacewires. Both catheters were from Cordis Corporation, Florida, USA and were inserted using Seldinger technique. A Foley catheter with a temperature gauge (Covidien, Degania Bet) was inserted into the bladder for urination and monitoring of temperature. To keep temperature constant during the experiment a forced air warming system was used (Mistral-Air Stryker, Portage, MI, USA). Access to the right jugular vein was done with surgical cut down.

## S3. Coordinate system transformation of implanted sensor data

The implanted motion sensor measures the 3D linear acceleration and angular velocity in the moving sensor's body frame (X, Y, Z), which follows the orientation of the sensor, as shown in Figure S1. The Z axis of the sensor's moving body frame is aligned with the normal vector to the local heart surface. However, the operation of proposed energy harvesters relates to inertial force that should be analyzed in the fixed reference coordinate system (x, y, z). Therefore, it is necessary to transform measured data from the body frame to the fixed reference frame. For this coordinate system transformation, the relative 3D rotation of the sensor throughout cardiac cycles should be considered, as shown in Figure S1.

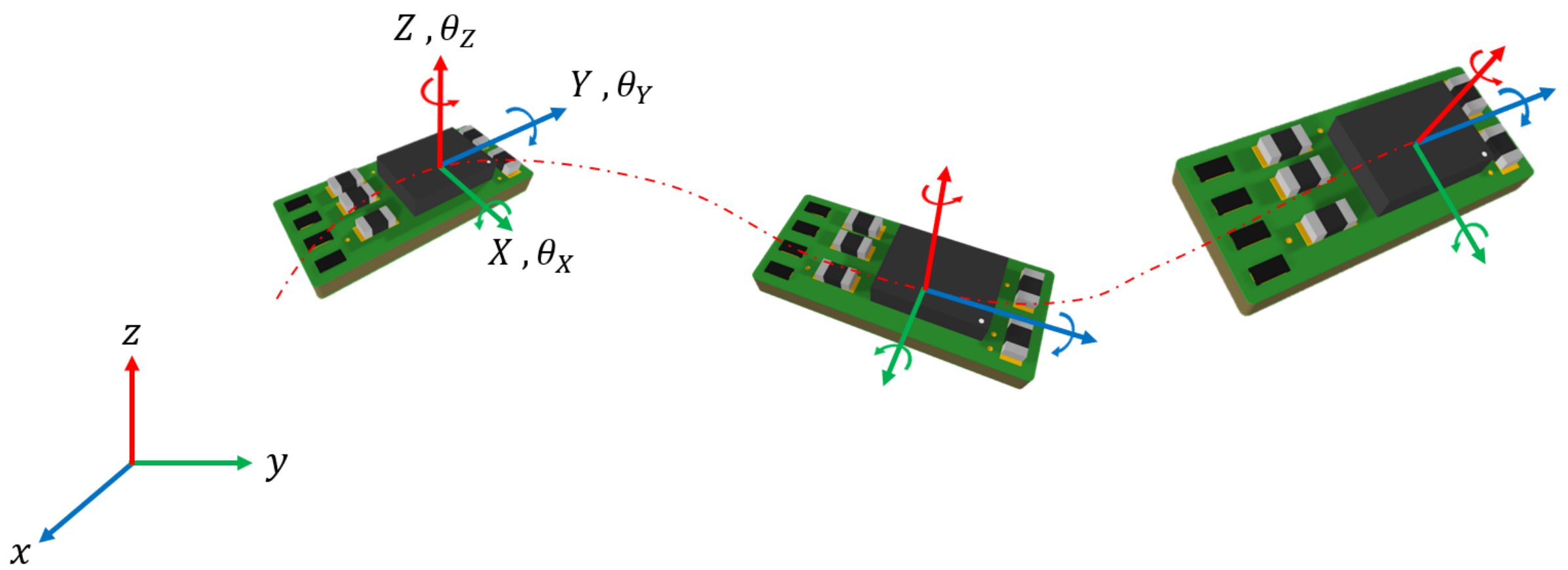

Figure S1- The variable orientation of the sensor's body frame with respect to fixed reference coordinates.

The built-in fusion algorithm provides the real-time orientation based on Euler angles and quaternions. While Euler angles offer an intuitive representation of rotations, they suffer from the well-known issue of gimbal lock. Quaternions provide a robust and efficient representation of rotations by avoiding gimbal lock.

A quaternion consists of four components: one scalar part ($\mathrm{w}$) and three vector components ($\mathrm{x}, \mathrm{y}, \mathrm{z}$). Together, these components define the quaternion as:

$$q = \mathrm{w} + \mathrm{x}i + \mathrm{y}j + \mathrm{z}k \tag{s1}$$

where i, j, and k are unit vectors used to represent the imaginary components of the quaternion. A rotation matrix R can be defined as

$$[R] = \begin{bmatrix} 1 - 2(\mathrm{y}^2 + \mathrm{z}^2) & 2(\mathrm{xy} - \mathrm{wz}) & 2(\mathrm{xz} + \mathrm{wy}) \\ 2(\mathrm{xy} + \mathrm{wz}) & 1 - 2(\mathrm{x}^2 + \mathrm{z}^2) & 2(\mathrm{yz} - \mathrm{wx}) \\ 2(\mathrm{xz} - \mathrm{wy}) & 2(\mathrm{yz} + \mathrm{wx}) & 1 - 2(\mathrm{x}^2 + \mathrm{y}^2) \end{bmatrix} \tag{s2}$$

This rotation matrix enables the transformation of vectors from the body frame $\vec{v} = [v_X, v_Y, v_Z]^T$ to the reference frame $\vec{v}' = R.\vec{v}$. Thus, the measured instantaneous linear acceleration vector $\vec{a} = [a_X, a_Y, a_Z]^T$ and angular velocity vector $\vec{\Omega} = [\Omega_X, \Omega_Y, \Omega_Z]^T$ in the body frame are transformed into the fixed reference frame as:

$$\vec{a'}(t) = \begin{bmatrix} a_x \\ a_y \\ a_z \end{bmatrix} = [R] \begin{bmatrix} a_X \\ a_Y \\ a_Z \end{bmatrix} \tag{s3}$$

$$\overrightarrow{\Omega'}(t) = \begin{bmatrix} \Omega_x \\ \Omega_y \\ \Omega_z \end{bmatrix} = [R] \begin{bmatrix} \Omega_X \\ \Omega_Y \\ \Omega_Z \end{bmatrix} \tag{s4}$$

Henceforth, all vectors in the fixed reference frame are denoted with a superscript $'$, and the time derivative is shown using a dot. The transformed vectors $\overrightarrow{a'}(t)$ and $\overrightarrow{\Omega'}(t)$ represent the motion of the BNO055 chip's location that has an offset $\vec{r}$ from the heart surface (implant site), as shown in Fig. 3-(a) of the main text. The direction of the vector $\vec{r}$ in the body frame is fixed in the Z direction. The base acceleration $\overrightarrow{a'}_{base}$ at the heart surface is recovered from:

$$\overrightarrow{a'}_{base}(t) = \overrightarrow{a'}(t) - \dot{\overrightarrow{\Omega'}}(t) \times \overrightarrow{r'}(t) - \overrightarrow{\Omega'}(t) \times (\overrightarrow{\Omega'}(t) \times \overrightarrow{r'}(t)) \tag{s5}$$

According to Fig. 3 of the main text, while the cardiac motion is measured from the epicardial surface during these animal tests, this data can be employed as an approximation of endocardial motion to analyze endocardial energy harvesters.

Physiological motion over the heart encompasses both respiratory and heartbeat movements [3]. This research concentrates solely on heartbeat motion, so the respiratory motion is filtered out to isolate heartbeat motion. Therefore, the linear acceleration and rotational velocity vectors at the implant site are approximately periodic and can be described by:

$$\begin{gathered} \overrightarrow{a'}_{base}(t) = \overrightarrow{a'}_{base}(t+T) \\ \overrightarrow{\Omega'}(t) = \overrightarrow{\Omega'}(t+T) \end{gathered} \tag{s6}$$

where $T$ represents the duration of a cardiac cycle. All measured cases in this study are examined to derive both time-dependent vectors $\overrightarrow{a'}_{base}(t)$ and $\overrightarrow{\Omega'}(t)$.

## S4. Analysis of kinematic criteria variability and statistical distribution

To perform the post-processing, the raw data from all recorded individual cardiac cycles was first loaded. We then established a set of 10-bpm heart-rate (HR) bins (e.g., 75-85 bpm, 85-95 bpm), and each individual cycle was programmatically assigned to one of these bins based on its measured heart rate. For each of the 12 kinematic criteria, the full collection of data points from all cycles within a given bin was analyzed statistically.

The observed variability within each bin stems from a combination of factors: the inherent physiological cycle-to-cycle variation (as no two heartbeats are mechanically identical), the superimposed motion from the respiratory cycle (as breathing motion can affect the measured heart motion), the 10-bpm bin width itself (which groups data from a range of heart rates), and minor sensor measurement noise. This statistical distribution was characterized using the box-and-whisker plots shown in the supplementary material (Section S1). In these plots, the central red line indicates the median value, while the "box" itself represents the interquartile range (IQR). The notches on the box visually represent the 95% confidence interval for the median. The "whiskers" extend to the most extreme data points not considered outliers, and individual outliers are plotted as separate red '+' symbols. This approach allowed us to robustly quantify this combined variability and compare it against the median performance.

- **Implant site 1 (Mid-Septum right):**

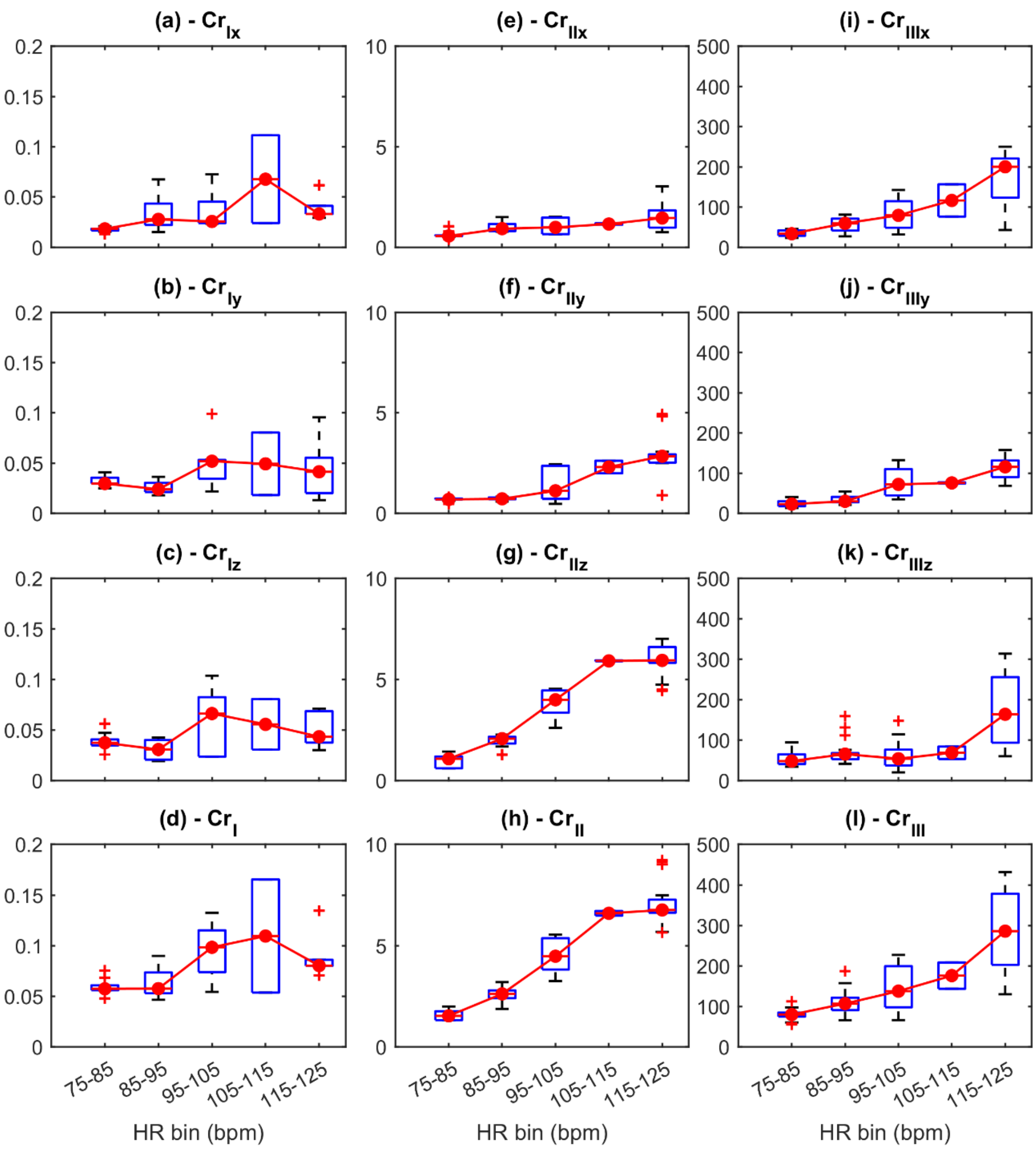

*The figures in the first, second, and third columns are plotted in SI units: m/s, m/s[2], and m/s[3], respectively.

Figure S3- The criteria values at the implant site 1 (mid-septum right); kinetic energy criteria: (a) $Cr_{Ix}$, (b) $Cr_{Iy}$, (c) $Cr_{Iz}$, (d) $Cr_{I}$, acceleration criteria: (e) $Cr_{IIx}$, (f) $Cr_{IIy}$, (g) $Cr_{IIz}$, (h) $Cr_{II}$, and jerk criteria: (i) $Cr_{IIIx}$, (j) $Cr_{IIIy}$, (k) $Cr_{IIIz}$, (l) $Cr_{III}$.

- **Implant site 2 (Right ventricle – outflow tract):**

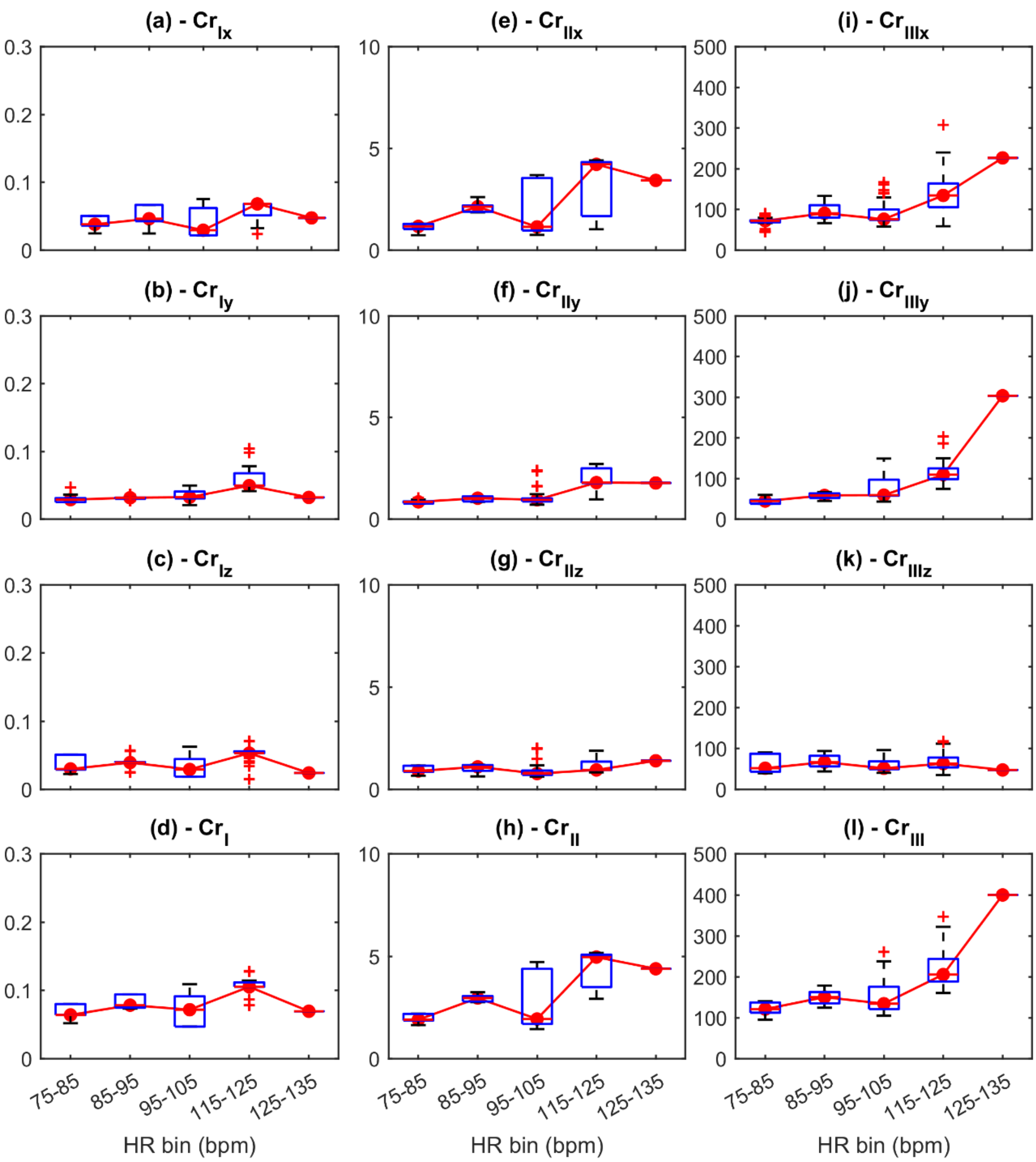

*The figures in the first, second, and third columns are plotted in SI units: m/s, m/s$^2$, and m/s$^3$, respectively.

Figure S4- The criteria values at the implant site 2 (right ventricle – outflow tract); kinetic energy criteria: (a) $Cr_{Ix}$, (b) $Cr_{Iy}$, (c) $Cr_{Iz}$, (d) $Cr_{I}$, acceleration criteria: (e) $Cr_{IIx}$, (f) $Cr_{IIy}$, (g) $Cr_{IIz}$, (h) $Cr_{II}$, and jerk criteria: (i) $Cr_{IIIx}$, (j) $Cr_{IIIy}$, (k) $Cr_{IIIz}$, (l) $Cr_{III}$.

- **Implant site 3 (Basal lateral):**

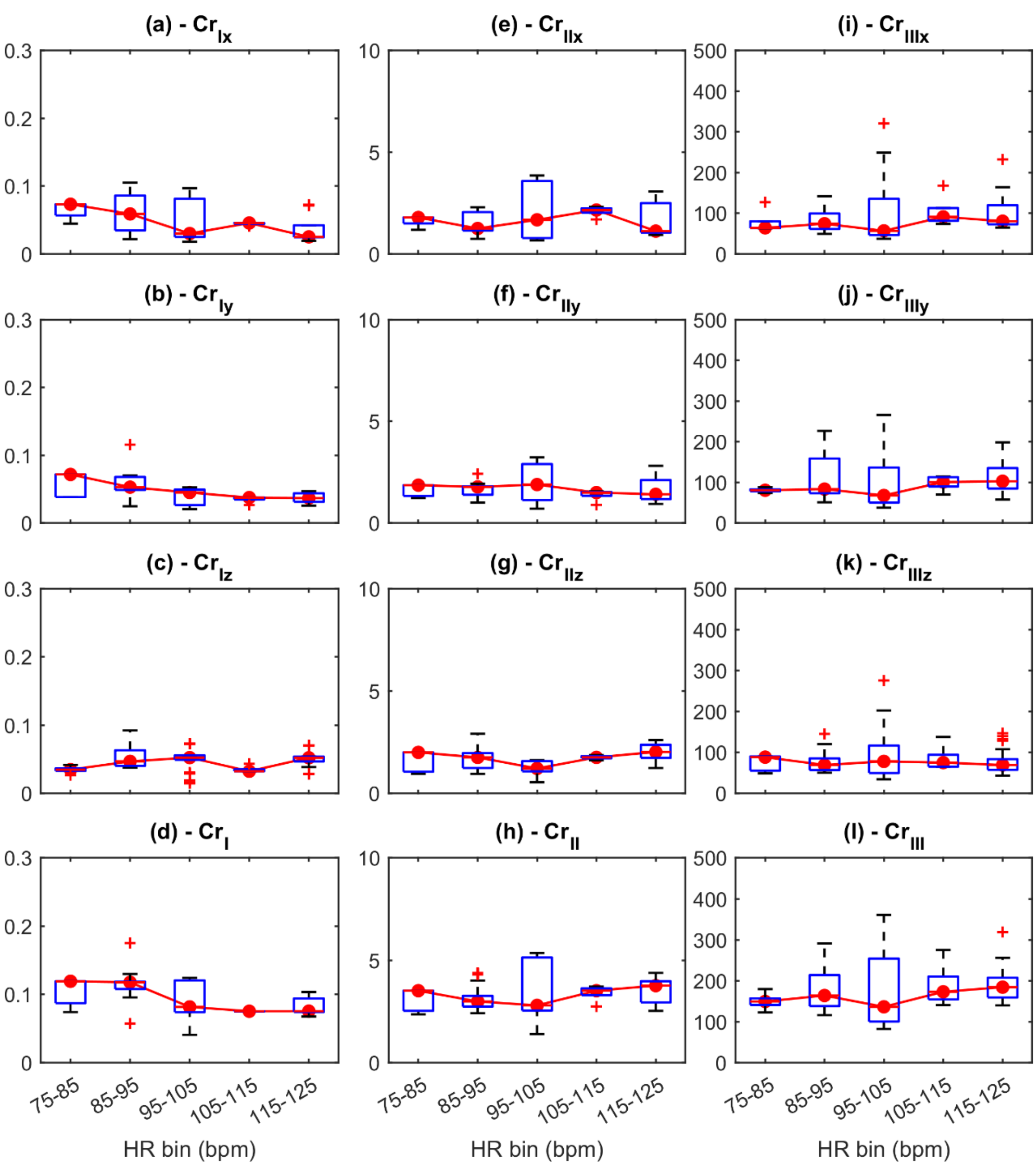

*The figures in the first, second, and third columns are plotted in SI units: m/s, m/s$^2$, and m/s$^3$, respectively.

Figure S5- The criteria values at the implant site 3 (basal lateral); kinetic energy criteria: (a) $Cr_{Ix}$, (b) $Cr_{Iy}$, (c) $Cr_{Iz}$, (d) $Cr_{I}$, acceleration criteria: (e) $Cr_{IIx}$, (f) $Cr_{IIy}$, (g) $Cr_{IIz}$, (h) $Cr_{II}$, and jerk criteria: (i) $Cr_{IIIx}$, (j) $Cr_{IIIy}$, (k) $Cr_{IIIz}$, (l) $Cr_{III}$.

- **Implant site 4 (High septum):**

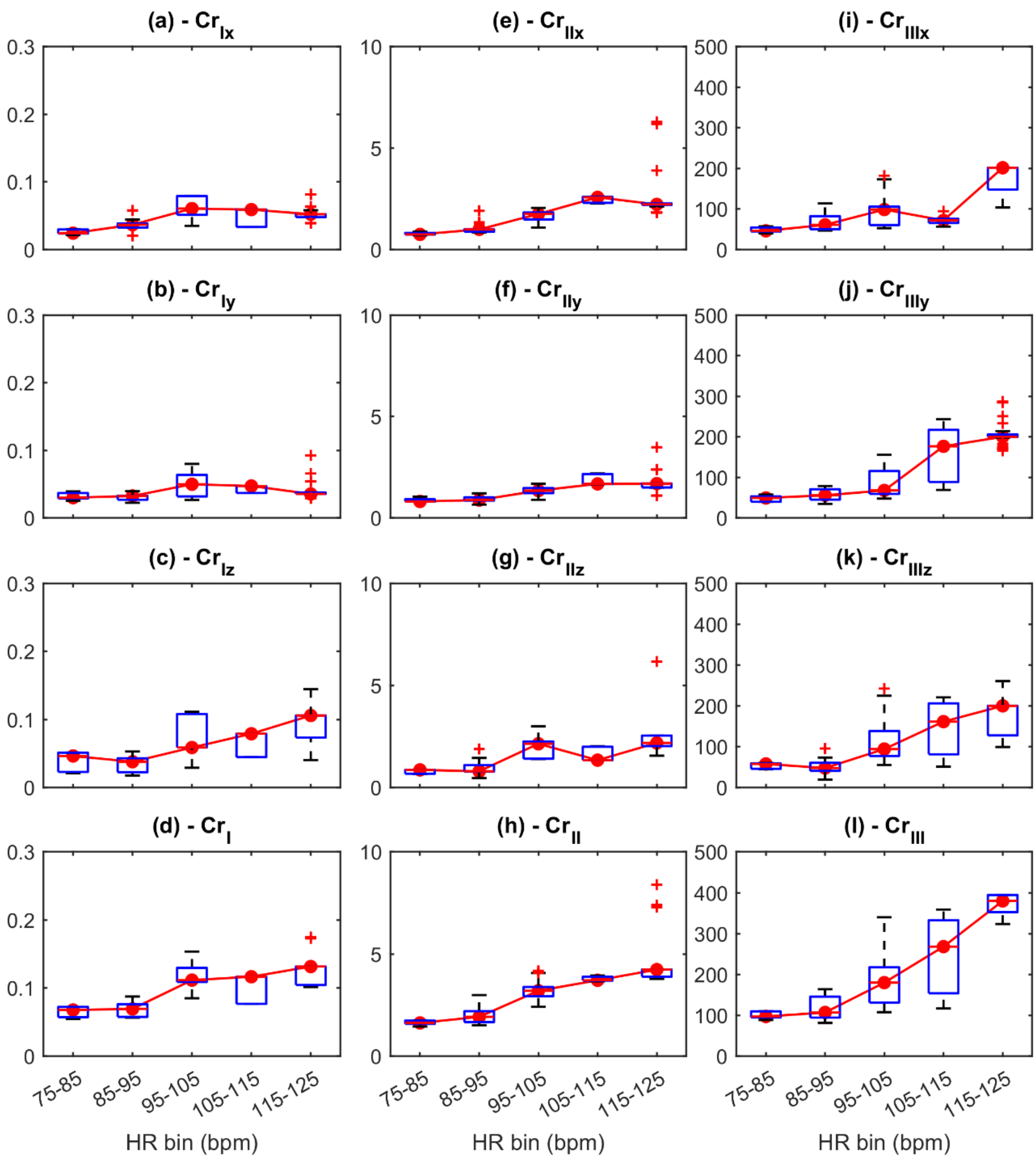

*The figures in the first, second, and third columns are plotted in SI units: m/s, m/s$^2$, and m/s$^3$, respectively.

Figure S6- The criteria values at the implant site 4 (high septum); kinetic energy criteria: (a) $Cr_{Ix}$, (b) $Cr_{Iy}$, (c) $Cr_{Iz}$, (d) $Cr_{I}$, acceleration criteria: (e) $Cr_{IIx}$, (f) $Cr_{IIy}$, (g) $Cr_{IIz}$, (h) $Cr_{II}$, and jerk criteria: (i) $Cr_{IIIx}$, (j) $Cr_{IIIy}$, (k) $Cr_{IIIz}$, (l) $Cr_{III}$.

- **Implant site 5 (Mid-anterior right ventricle):**

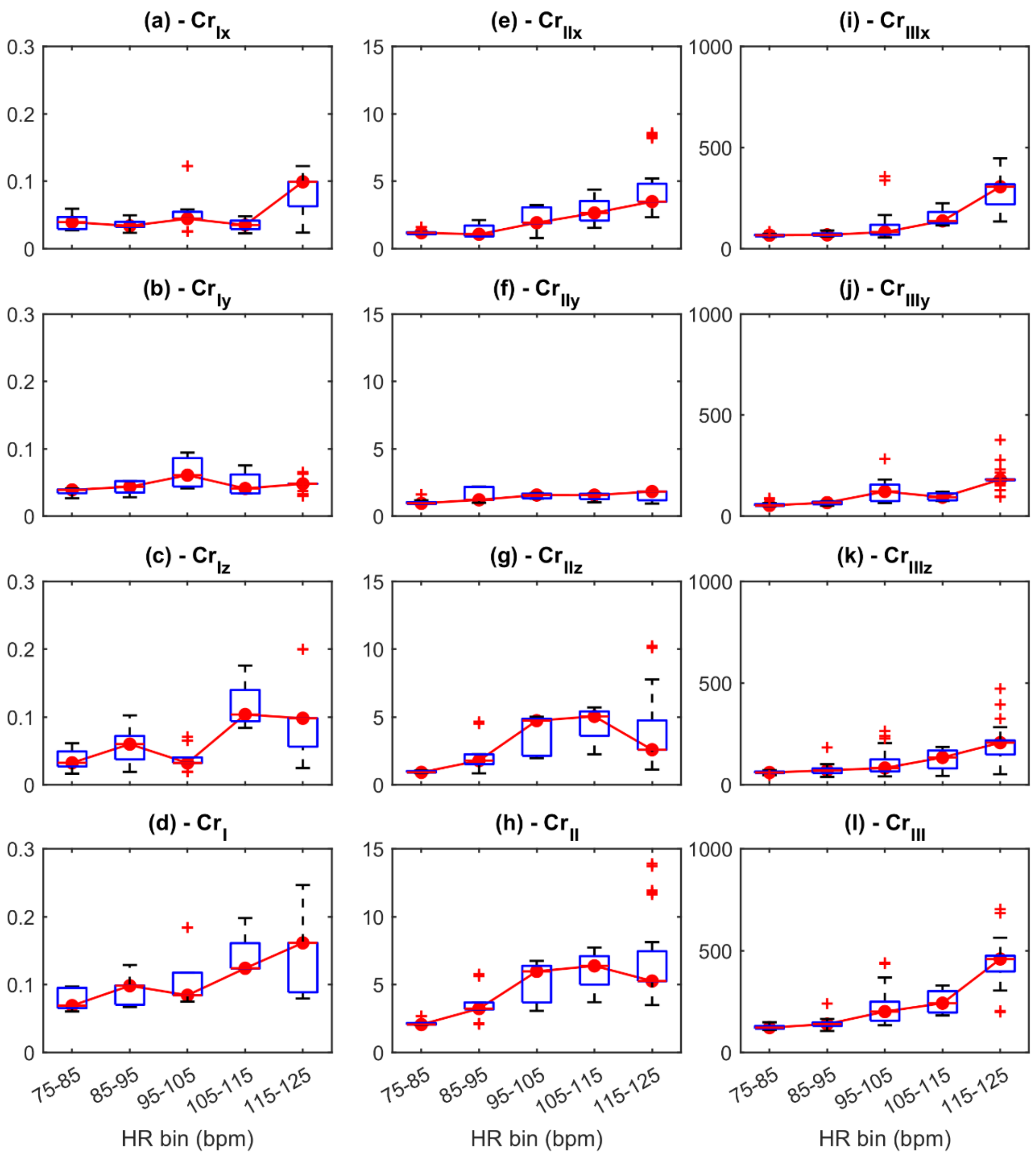

*The figures in the first, second, and third columns are plotted in SI units: m/s, m/s$^2$, and m/s$^3$, respectively.

Figure S7- The criteria values at the implant site 5 (mid-anterior right ventricle); kinetic energy criteria: (a) $Cr_{Ix}$, (b) $Cr_{Iy}$, (c) $Cr_{Iz}$, (d) $Cr_{I}$, acceleration criteria: (e) $Cr_{IIx}$, (f) $Cr_{IIy}$, (g) $Cr_{IIz}$, (h) $Cr_{II}$, and jerk criteria: (i) $Cr_{IIIx}$, (j) $Cr_{IIIy}$, (k) $Cr_{IIIz}$, (l) $Cr_{III}$.

- **Implant site 6 (Apex, left ventricle):**

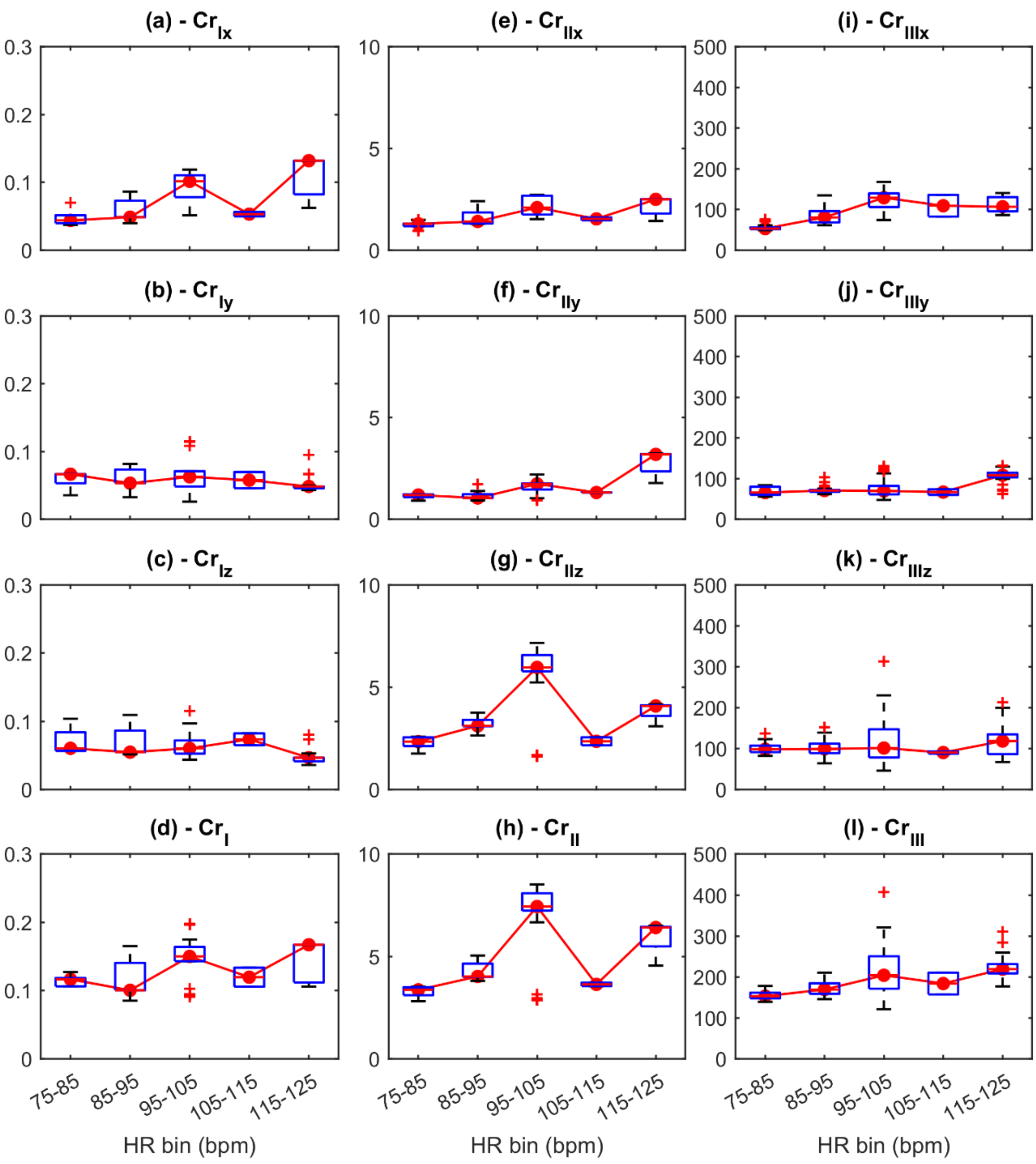

*The figures in the first, second, and third columns are plotted in SI units: m/s, m/s$^2$, and m/s$^3$, respectively.

Figure S8- The criteria values at the implant site 6 (apex, left ventricle); kinetic energy criteria: (a) $Cr_{Ix}$, (b) $Cr_{Iy}$, (c) $Cr_{Iz}$, (d) $Cr_{I}$, acceleration criteria: (e) $Cr_{IIx}$, (f) $Cr_{IIy}$, (g) $Cr_{IIz}$, (h) $Cr_{II}$, and jerk criteria: (i) $Cr_{IIIx}$, (j) $Cr_{IIIy}$, (k) $Cr_{IIIz}$, (l) $Cr_{III}$.

## S5. The effect of rotational motion in power generation

The considered energy harvester consists of 25 piezoelectric beams. Figure S2 presents the instantaneous power generation of beam 15 with and without considering rotational motion at a specific motion measurement case. The results indicate that rotational motion can affect the energy harvesting level of each piezoelectric beam.

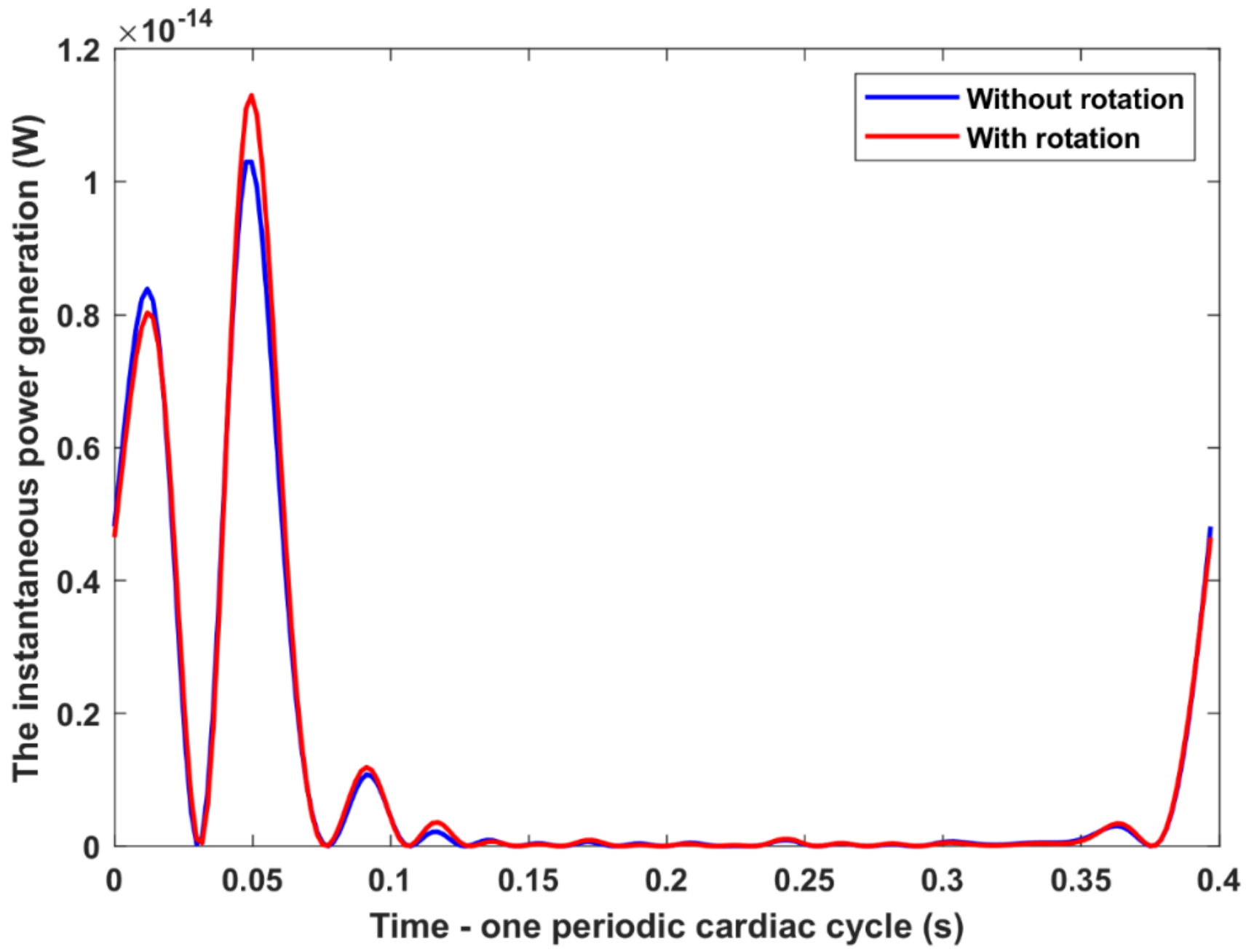

Figure S9- The comparison of the instantaneous power generated by the single piezoelectric beam 15 with and without considering rotational motion at a specific motion measurement case.